\documentclass{article}
\usepackage[utf8]{inputenc}

\usepackage{amssymb}
\usepackage{graphicx}
\usepackage{authblk}
\usepackage[utf8]{inputenc}
\usepackage{mathptmx}
\usepackage{amsmath}
\usepackage{epsfig,amsopn}
\usepackage{graphicx}
\usepackage{braket}
\usepackage{float}
\usepackage{enumerate}
\usepackage[normalem]{ulem}
\usepackage{todonotes}
\newcommand{\eref}[1]{Eq.~(\ref{#1})}%
\DeclareMathOperator\erf{erf}

\begin{document}

\title{Mean-performance of Sharp Restart II:\\
Inequality Roadmap}
\author[1]{Iddo Eliazar \thanks{email: eliazar@tauex.tau.ac.il}}
\author[1]{Shlomi Reuveni \thanks{email: shlomire@tauex.tau.ac.il}}
\affil[1]{School of Chemistry, Center for the Physics and Chemistry of Living Systems, The Sackler Center for Computational Molecular and Materials Science, and The Mark Ratner Institute for Single Molecule Chemistry, Tel Aviv University, Tel Aviv 6997801, Israel}

\maketitle

\begin{abstract}

Restarting a deterministic process always impedes its completion. However, it is known that restarting a random process can also lead to an opposite outcome -- expediting completion. Hence, the effect of restart is contingent on the underlying statistical heterogeneity of the process' completion times. To quantify this heterogeneity we bring a novel approach to restart: the methodology of inequality indices, which is widely applied in economics and in the social sciences to measure income and wealth disparity. Using this approach we establish an `inequality roadmap' for the mean-performance of sharp restart: a whole new set of universal inequality criteria that determine when restart with sharp timers (i.e. with fixed deterministic timers) decreases/increases mean completion. The criteria are based on a host of inequality indices including Bonferroni, Gini, Pietra, and other Lorenz-curve indices; each index captures a different angle of the restart-inequality interplay. Utilizing the fact that sharp restart can match the mean-performance of any general restart protocol, we prove -- with unprecedented precision and resolution -- the validity of the following statement: restart impedes/expedites mean completion when the underlying statistical heterogeneity is low/high.

\bigskip\ 


\bigskip\ 

\end{abstract}

\newpage

\section{\label{1}Introduction}

This paper is the second in a duo of works exploring the mean-performance of sharp restart. In part one we presented a comprehensive statistical analysis of the topic \cite{MP1SR}. In this part we continue the exploration, doing so from the rather surprising perspective of socioeconomic inequality. Indeed, this paper shall reveal profound connections between two seemingly unrelated topics: the measurement of socioeconomic inequality on the one hand, and the mean-performance of sharp restart on the other hand. The novel inequality vantage point yields a whole new set of universal inequality criteria that pinpoint -- with unprecedented precision and resolution -- how the mean-performance of general restart protocols is contingent on the underlying statistical heterogeneity.

As explained in part one, sharp restart can be perceived as an algorithm. The algorithm's input is the random duration of a general task to be accomplished -- e.g. a completion time or a first-passage time of a given stochastic process \cite{gardiner}-\cite{zilman}. As long as the task is not accomplished, the algorithm restarts the stochastic process periodically, using a `sharp' timer -- i.e. a fixed deterministic timer. The algorithm's output is the random duration it takes to accomplish the task `under restart'. The motivation for using sharp restart, together with a rather rich literature survey (including \cite{Restart1}-\cite{Restart21}), was described in detail in part one \cite{MP1SR}.

Following a main path in restart research, the analysis presented in \cite{MP1SR} focused on mean performance: comparing the input's mean to the output's mean. Sharp restart was termed ``beneficial'' if the output's mean is smaller than the input's mean -- i.e. if mean completion is expedited. Conversely, sharp restart was termed ``detrimental'' if the output's mean is larger than the input's mean -- i.e. if mean completion is impeded. The two principal analytic tools employed in \cite{MP1SR} were the residual lifetime of renewal theory \cite{Smi}-\cite{Cin}, and the hazard rate of reliability engineering \cite{BP}-\cite{Dhi}. Using these tools, we established a detailed statistical roadmap for the mean-performance of sharp-restart: universal statistical criteria that determine when sharp restart is beneficial, and when it is detrimental.

A key result in \cite{MP1SR} was based on the input's coefficient of variation (CV), the ratio between the standard deviation and the mean of the task's duration. The CV is a principal standardized measure of statistical heterogeneity. Using this measure we established the following threshold scheme: if the input's CV is larger than one then there exist timers with which sharp restart is beneficial; and if the input's CV is smaller than one then there exist timers with which sharp restart is detrimental.\footnote{The CV threshold scheme holds when the input's mean is finite. If the input's mean is infinite then sharp restart is always beneficial.} A similar CV result holds with regard to the introduction of exponential restart -- in which the periods between consecutive restart epochs are drawn (independently) from the exponential distribution \cite{FPUR1}-\cite{CV7}. Moreover, when adding branching to exponential restart, the input's Gini coefficient (to be explained below) enters the picture and joins the input's CV in determining if the introduction of exponential restart is detrimental or beneficial \cite{Branching}.

In economics, the Gini coefficient is a principal inequality index \cite{Gin1}-\cite{GG2}. Widely applied in the social sciences, inequality indices are quantitative gauges of socioeconomic disparity \cite{Cou}-\cite{Cow}. Namely, given a human society of interest, inequality indices score the disparity of income and wealth among the society's members. While devised to gauge socioeconomic disparity, the use of inequality indices is not at all limited to this particular context. Indeed, inequality indices can be used in a far broader context: measuring the statistical heterogeneity of general non-negative random variables \cite{StaEve}-\cite{TouInq}. The CV measures statistical heterogeneity from an Euclidean-geometry perspective, inequality indices do so from a socioeconomic-inequality perspective, and these perspectives are different \cite{HarInq}.

Here we apply inequality indices to measure  the statistical heterogeneity of the task's random duration. Taking this approach we establish a whole new set of universal inequality criteria that determine when sharp restart is beneficial, and when it is detrimental. All the inequality criteria share a common `bedrock structure' that is described by the following take-home-message statement: high statistical heterogeneity implies that sharp restart is beneficial; and low statistical heterogeneity implies that sharp restart is detrimental. Using precise threshold levels, the inequality criteria articulate this statement via exact and explicit mathematical results -- thus providing an `inequality roadmap' for the mean-performance sharp restart. In turn, as sharp restart can match the mean-performance of any other restart protocol \cite{FPUR1,FPUR2}, the take-home-message statement holds for general restart protocols.


We begin with three inequality indices: a CV-based index, and the Gini and Bonferroni indices.\footnote{The terms ``Gini coefficient'' and ``Gini index'' are used interchangeably in the literature.} With regard to these indices we establish existence criteria similar to that of the aforementioned CV result \cite{MP1SR}. Namely, each index has a corresponding threshold level, and: if the index is larger/smaller than its threshold level then there exist timers with which sharp restart is beneficial/detrimental. These three existence criteria are different, as their indices capture different `angles' of inequality. Consequently, each of the existence criteria provides a different type of information regarding the restart-inequality interplay.

While existence criteria are important, they cannot assert if sharp restart with a specific timer is beneficial or detrimental. For practical uses, it is evident that timer-specific criteria are essential. This practical need leads to a profound theoretical question: can inequality provide timer-specific information? To that end, we first focus on two sharp-restart timers of special interest -- one that equals the input's mean, and one that equals the input's median. We show that two inequality indices, the Pietra index and vertical-diameter (Vdiam) index, relate to these timers respectively. With regard to these indices we establish timer-specific criteria for the mean and median timers. Namely, each index has a corresponding threshold level, and: if the index is larger/smaller than its threshold level then sharp restart with the compatible timer is beneficial/detrimental.

From the mean and median timers we carry on to general timers. To that end, we consider two continuums of vertical and horizontal Lorentz-curve inequality indices. Given a general timer, we show that this timer is related to a unique index of the vertical continuum, as well as to a unique index of the horizontal continuum. With regard to these unique indices we establish timer-specific criteria. Namely, each unique index has a corresponding threshold level, and: if the index is larger/smaller than its threshold level then sharp restart with the given timer is beneficial/detrimental. Special cases of these general results include the aforementioned Pietra and Vdiam criteria, as well as a `horizontal counterpart' of the Vdiam criteria.

The remainder of this paper is organized as follows. Section \ref{2} sets the stage: it recaps sharp restart and the aforementioned CV result \cite{MP1SR}; and it reviews the notion of inequality indices, as well as the very foundation of many inequality indices -- the Lorentz curve. Sections \ref{3}-\ref{5} present, respectively: the Gini and Bonferroni existence criteria; the Pietra and Vdiam mean and median criteria; and the general criteria that are based on the vertical and horizontal continuums. Section \ref{7} concludes with an `inequality roadmap' for the mean-performance of sharp restart: a summary of the key results established here.

A note about notation: along the paper $\mathbf{E}\left[ \xi \right]$ denotes the expectation of a (non-negative) random variable $\xi $; and IID is acronym for independent and identically distributed (random variables). 

\section{\label{2}Setting the stage}

This section sets the stage for a comprehensive mean-performance analysis of sharp restart -- to be based on socioeconomic inequality indices -- that will be carried out in the next sections. Subsections \ref{20} and \ref{21} recap material that was presented in \cite{MP1SR}. Subsection \ref{22} and \ref{23} review, respectively, the socioeconomic notions of inequality indices and the Lorenz curve.

\subsection{\label{20}Sharp restart}

\emph{Sharp restart} admits the following algorithmic description. There is a general task with completion time $T$, a positive-valued random variable. To this task a three-steps algorithm, with a positive deterministic timer $\tau $, is applied. Step I: initiate simultaneously the task and the timer. Step II: if the task is accomplished up to the timer's expiration -- i.e. if $T\leq \tau $ -- then stop upon task completion. Step III: if the task is not accomplished up to the timer's expiration -- i.e. if $T>\tau $ -- then, as the timer expires, go back to Step I and start afresh (independently of the past).

The sharp-restart algorithm generates an iterative process of independent and statistically identical task-completion trials. This process halts during its first successful trial, and we denote by $T_{R}$ its halting time. Namely, $T_{R}$ is the overall time it takes -- when the sharp-restart algorithm is applied -- to complete the task. The sharp-restart algorithm is a \emph{non-linear mapping} whose input is the random variable $T$, whose output is the random variable $T_{R}$, and whose parameter is the
deterministic timer $\tau $.

This paper shall use the following notation regarding the input's statistics: distribution function, $F\left( t\right) =\Pr \left( T\leq t\right) $ ($t\geq 0$); survival function, $\bar{F}\left( t\right) =\Pr\left( T>t\right) $ ($t\geq 0$); density function, $f\left( t\right)=F^{\prime }\left( t\right) =-\bar{F}^{\prime }\left( t\right) $ ($t>0$); and mean, $\mu =\mathbf{E}\left[ T\right] =\int_{0}^{\infty }tf\left(t\right) dt$. The input's density function is henceforth considered to be positive-valued over the positive half-line: $f\left( t\right) >0$ for all $t>0$.\footnote{This is merely a technical assumption, which is introduced in order to assure that all positive timers $\tau$ are admissible. In general, admissible timers are in the range $t_{low} < \tau < \infty$, where $t_{low}$ is the lower bound of the support of the input's density function: $t_{low}=\inf \left\{ t>0 \:|\: f\left( t\right) >0\right\}$.}

Throughout this paper $M\left( \tau \right) =\mathbf{E}\left[ T_{R}\right] $ shall denote the output's mean; this notation underscores the fact that the output's mean is a function of the timer $\tau $, the parameter of the sharp-restart algorithm. In terms of the input's distribution and survival functions, the output's mean is given by \cite{MP1SR,BP,FPUR1}: 
\begin{equation}
M\left( \tau \right) =\frac{1}{F\left( \tau \right) }\int_{0}^{\tau }\bar{F}\left( t\right) dt.  \label{211}
\end{equation}
As the input's mean is the integral of the input's survival function, $\mu =\int_{0}^{\infty }\bar{F}\left( t\right) dt$, Eq. (\ref{211}) implies that: in the limit $\tau \rightarrow \infty $, the output's mean coincides with the input's mean, $\lim_{\tau \rightarrow \infty }M\left( \tau \right) =\mu $.

Examining the sharp-restart algorithm from a mean-performance perspective, it is key to determine if the application of the algorithm will expedite task-completion, or if it will impede task-completion. To that end the following terminology shall be used \cite{MP1SR}:

\begin{enumerate}
\item[$\bullet $] Sharp restart with timer $\tau $ is \emph{beneficial} if it improves mean-performance, $M\left( \tau \right) <\mu $.

\item[$\bullet $] Sharp restart with timer $\tau $ is \emph{detrimental} if it worsens mean-performance, $M\left( \tau \right) >\mu $.
\end{enumerate}

\noindent Eq. (\ref{211}) implies that the output's mean is always finite: $M\left(\tau \right) <\infty $ for all timers $\tau $. Thus, if the input's mean is infinite, $\mu =\infty $, then the application of the sharp-restart algorithm is highly beneficial -- as it reduces the input's infinite mean to the output's finite mean: $M\left( \tau \right) <\mu =\infty $. Having resolved the case of infinite-mean inputs, we henceforth set the focus on the case of positive-mean inputs, $0< \mu <\infty $.

\subsection{\label{21}CV criteria}

The variance of the input $T$ is the input's mean square deviation from its mean, $\sigma ^{2}=\mathbf{E}[\left\vert T-\mu \right\vert ^{2}]$. The input's standard deviation $\sigma $ is the square root of its variance. And, the input's coefficient of variation (CV) is the ratio of its standard deviation to its mean, $\sigma / \mu  $. The input's CV is a `normalized' version of its standard deviation $\sigma$. 

In the first part of this duo, based on the input's CV, the following pair of \emph{CV criteria} was established \cite{MP1SR}:

\begin{enumerate}
\item[$\bullet $] If the CV is smaller than one -- which is equivalent to $\sigma < \mu$ -- then  there exist timers $\tau $ for which sharp restart is detrimental.

\item[$\bullet $] If the CV is larger than one -- which is equivalent to $\sigma > \mu$ -- then there exist timers $\tau $ for which sharp restart is beneficial.
\end{enumerate} 

\noindent The CV criteria stem from the following formula: 

\begin{equation}
\int_{0}^{\infty }\left[ M\left( \tau \right) -\mu \right] F\left( \tau \right) d\tau =\frac{1}{2}\left( \mu ^{2}-\sigma ^{2}\right).
\label{212}
\end{equation}

\noindent Namely, Eq. (\ref{212}) manifests how the interplay between the input's standard deviation and mean -- $\sigma $ versus $\mu$ -- affects the output's mean $M\left( \tau \right) $. Equivalently, Eq. (\ref{212}) manifests the effect of the input's CV on the output's mean $M\left( \tau \right) $. The derivation of Eq. (\ref{212}) is detailed in the Methods.

The CV criteria have the following threshold form. The input's CV is compared to the threshold level $1$. To determine the existence of timers for which the application of the sharp-restart algorithm is detrimental or beneficial, one needs to check if the input's CV is below or above the threshold level, respectively. Evidently, in the CV criteria, one can replace the CV by its square, $ \sigma^{2} / \mu^{2} $.  

\subsection{\label{22}Inequality indices}

Consider a human society comprising of members with non-negative wealth values. Such a society has two socioeconomic extremes: perfect equality and perfect inequality. In a perfectly equal society all members share a common positive wealth value. In a perfectly unequal society $0\%$ of the members possess $100\%$ of the overall wealth; in this socioeconomic extreme the society's population is assumed infinitely large \cite{HarInq}.

An \emph{inequality index} $\mathcal{I}$ is a measure of the society's socioeconomic inequality: the larger the index -- the more socioeconomically unequal the society. More specifically, an inequality index $\mathcal{I}$ takes values in the unit interval, $0\leq \mathcal{I}\leq 1$, and it has three basic properties \cite{Cou}-\cite{Cow}. (I) The index meets its zero lower bound $\mathcal{I}=0$ if and only if the society is perfectly equal. (II) If the society is perfectly unequal then the index meets its unit upper bound $\mathcal{I}=1$. (III) The index is invariant with respect to the currency via which wealth is measured.

While widely used in economics and in the social sciences \cite{Cou}-\cite{Cow}, the application of inequality indices is not confined to these fields alone. Indeed, an inequality index $\mathcal{I}$ can be used to measure the inherent `socioeconomic inequality' of any given non-negative random variable with a finite mean \cite{HarInq,TouInq}. To that end, deem the random variable under consideration to represent the wealth of a randomly-sampled member of a virtual society. Then, the socioeconomic inequality of the random variable is that of its corresponding virtual society. 

Henceforth, the random variable under consideration is the input $T$ of the sharp-restart algorithm. An illustrative example of an inequality index of the input $T$ is what shall be referred to here as \emph{CV index}. This inequality index was introduced in \cite{RenEq}, and it is a special case of the input's R\'{e}nyi spectra \cite{TouInq}-\cite{RenEq}. In terms of the input's CV, $\sigma / \mu$, the input's CV index admits the following representation \cite{RenEq}:

\begin{equation}
\mathcal{I}_{CV}=1-\frac{1}{1+ (\sigma / \mu)^{2}}.
\label{220}
\end{equation}

The input's CV index $\mathcal{I}_{CV}$ is a monotone increasing function of the input's CV. This index meets its zero lower bound if and only if the input's CV vanishes, i.e. -- for a given positive-mean input -- if and only if the input's standard deviation vanishes, $\sigma=0$. And, this index meets its unit upper bound if and only if the input's CV diverges, i.e. -- for a given positive-mean input -- if and only if the input's standard deviation diverges, $\sigma=\infty$. For a more detailed account of the CV index and its properties, see \cite{TouInq}. 

In terms of the input's CV index $\mathcal{I}_{CV}$, the pair of the CV criteria of subsection \ref{21} can be re-formulated as the following pair of \emph{CV-index criteria}:

\begin{enumerate}
\item[$\bullet $] If the input's CV index is smaller than half, $\mathcal{I}_{CV}<\frac{1}{2}$, then there exist timers $\tau $ for which sharp restart is detrimental.

\item[$\bullet $] If the input's CV index is larger than half, $\mathcal{I}_{CV}>\frac{1}{2}$, then there exist timers $\tau $ for which sharp restart is beneficial.
\end{enumerate}

The threshold form of the CV criteria is induced to the CV-index criteria. Specifically, the input's CV index $\mathcal{I}_{CV}$ -- the input's `CV inequality' -- is compared to the threshold level $\frac{1}{2}$. To determine the existence of timers for which the application of the sharp-restart algorithm is detrimental or beneficial, one needs to check if the input's `CV inequality' $\mathcal{I}_{CV}$ is below or above the threshold level $\frac{1}{2}$, respectively.

\subsection{\label{23} The Lorenz curve}

The CV-index criteria described above expose a connection between sharp restart and the measurement of socioeconomic inequality. In the following sections we shall show that this connection extends far beyond the CV-index criteria. To that end we shall use an additional socioeconomic working tool -- the Lorenz curve -- which is described in this subsection. Armed with the Lorenz curve, we shall show that the connection between sharp restart and inequality indices is profound and broad, and that inequality indices purvey valuable information on whether sharp restart is detrimental or beneficial.

As in subsection \ref{22}, consider a human society comprising of members with non-negative wealth values. The distribution of wealth among the society's members is quantified by the society's \emph{Lorenz curve} $y=L\left( x\right) $ ($0\leq x,y\leq 1$) \cite{Lor}-\cite{Cho}. Specifically, the Lorenz curve $y=L\left( x\right) $ has the following socioeconomic meaning: the low (poor) $100x\%$ of the society members possess $100y\%$ of the society's overall wealth.

The Lorenz curve $y=L\left( x\right) $ resides in the unit square ($0\leq x,y\leq 1$), and it has three basic properties \cite{Lor}-\cite{Cho}. (I) It `starts' at the square's bottom-left corner, $L\left( 0\right) =0$; and it `ends' at the square's top-right corner, $L\left( 1\right) =1$. (II) It is monotone increasing and concave. (III) It is invariant with respect to the currency via which wealth is measured. The first two properties imply that the Lorenz curve $y=L\left( x\right) $ is bounded from above by the square's diagonal line $y=x$ (see Fig. 1).

\begin{figure}[t!]
\centering
\includegraphics[width=9cm]{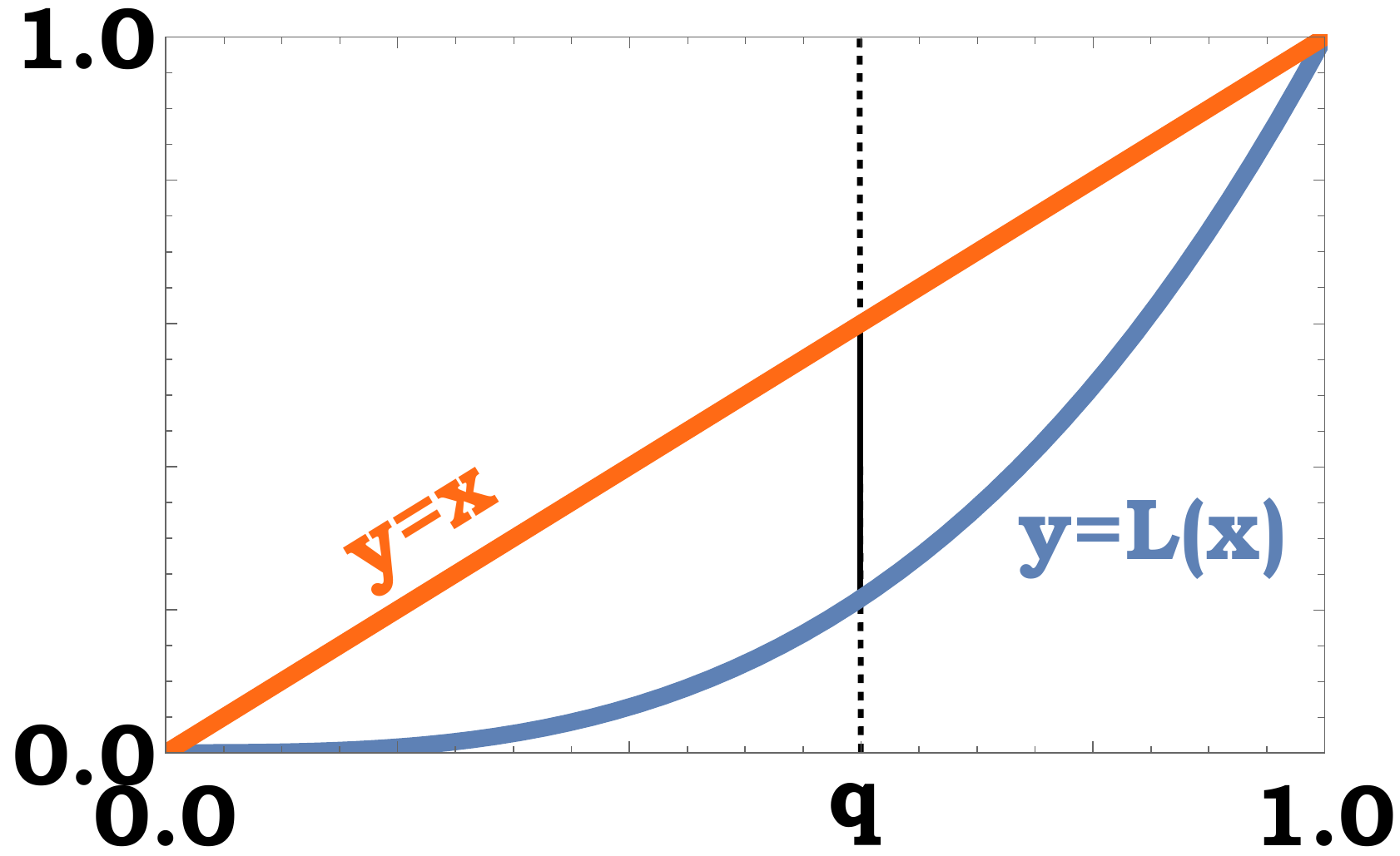}
\caption{Lorenz-curve illustration. The Lorenz curve $y=L(x)$ is depicted in blue, and the diagonal line $y=x$ is depicted in orange; the Lorenz curve and the diagonal line reside in the unit square ($0\leq x,y\leq 1$). The line segment depicted in solid black is the vertical distance -- along the vertical line $x=q$, depicted in dashed black -- between the Lorenz curve and the diagonal line.} 
\label{Lorenz curve}
\end{figure}

As noted in subsection \ref{22}, the human society under consideration has two socioeconomic extremes: perfect equality and perfect inequality. Recall that in a perfectly equal society all members share a common positive wealth value. In the space of Lorenz curves the perfect-equality socioeconomic extreme is characterized by the diagonal line $y=x$ ($0\leq x,y\leq 1$). Consequently, the deviation of the Lorenz curve $y=L\left( x\right) $ from the diagonal line $y=x$ can serve as a geometric gauge of the society's socioeconomic inequality: the deviation of the society from the perfect-equality socioeconomic extreme.

Evidently, there are many different ways of measuring the deviation of the Lorenz curve $y=L\left( x\right) $ from the diagonal line $y=x$. In turn, these different ways yield different inequality indices. An illustrative example is described as follows. For a fixed number $q$ (where $0<q<1$), consider the the vertical line $x=q$ of the unit square. The vertical distance -- along the vertical line $x=q$ -- between the Lorenz curve $y=L\left( x\right) $ and the diagonal line $y=x$ is: $q-L\left( q\right) $ (see Fig. 1). This vertical distance takes values in the range $\left[ 0,q\right] $. Consequently, the corresponding `normalized' vertical distance $[q-L\left(q\right) ]/q$ takes values in the unit interval $\left[ 0,1\right] $. It is straightforward to check that the normalized vertical distance $[q-L\left(q\right) ]/q$ meets the three inequality-index properties that were postulated in subsection \ref{22}.

Any non-negative valued random variable, with a positive mean, has a corresponding Lorenz curve \cite{HarInq,TouInq}. Indeed, as with inequality indices: deem the random variable under consideration to represent the wealth of a randomly-sampled member of a virtual society. Then, the Lorenz curve of the random variable is that of its corresponding virtual society. As above, we take the random variable under consideration to be the input $T$ of the sharp-restart algorithm. Thus, henceforth, $y=L\left( x\right) $ shall manifest the input's Lorenz curve. 

\section{\label{3}Gini and Bonferroni indices}

In this section we establish connections between sharp restart on the one hand, and the Gini and Bonferroni inequality indices on the other hand. Specifically, we shall show that these two inequality indices of the input $T$ yield existence criteria that are similar to the CV-index criteria of subsection \ref{22}, and that have a similar threshold form.

\subsection{\label{31}Gini-index criteria}

Perhaps the best-known and most popular socioeconomic inequality index is the \emph{Gini index} \cite{Gin1}-\cite{GG2}. This subsection addresses the Gini index $\mathcal{I}_{Gini}$ of the input $T$. 

In terms of the input's Lorenz curve, the input's Gini index $\mathcal{I}_{Gini}$ is twice the area captured between the Lorenz curve $y=L\left(x\right) $ and the diagonal line $y=x$ \cite{YS}. Namely,
\begin{equation}
\mathcal{I}_{Gini}=2\int_{0}^{1}\left[ q-L\left( q\right) \right] dq.
\label{611}
\end{equation}
The integral appearing on the right-hand side of Eq. (\ref{611}) is the average of the vertical distances $\left[ q-L\left( q\right) \right] $ ($0<q<1$) between the Lorenz curve and the diagonal line (see Fig. 1); these vertical distances were noted in subsection \ref{23}. 

The Gini-index representation of \eref{611} exhibits no clear connection to Eq. (\ref{211}), the output's mean. To make the connection more apparent we use the following, alternative, Gini-index representation \cite{TouInq}:

\begin{equation}
\mathcal{I}_{Gini}=1-\frac{1}{\mu }\mathbf{E}\left[ \min \left\{T_{1},T_{2}\right\} \right] ,  \label{311}
\end{equation}
where $T_{1}$ and $T_{2}$ are IID copies of the input $T$. The Gini-index representation of Eq. (\ref{311}) is based on the disparity between two means: the mean of the minimum $\min \{ T_{1},T_{2} \}$ versus the input's mean $\mu$. The smaller the disparity between the two means -- the closer is the Gini index to its zero lower bound. Conversely, the larger the disparity between the two means -- the closer is the Gini index to its unit upper bound.

In addition to the Gini-index representation of \eref{311}, we also use the following representation of the output's mean:

\begin{equation}
M\left( \tau \right) =\frac{1}{F\left( \tau \right) }\mathbf{E}\left[ \min \left\{ T,\tau \right\} \right].  \label{312} 
\end{equation} 
The derivation of Eq. (\ref{312}) is detailed in the Methods. Eqs. (\ref{311}) and (\ref{312}) incorporate similar terms -- the term $\mathbf{E}\left[ \min \left\{ T_{1},T_{2}\right\} \right] $ in the former, and the term $\mathbf{E}\left[ \min \left\{ T,\tau \right\} \right] $ in the latter. This similarity suggests that a connection between the input's Gini index $\mathcal{I}_{Gini}$ and the output's mean $M\left( \tau \right) $ may exist. We shall now establish such a connection.

Denote by $f_{\max }\left( t\right) $ ($t\geq 0$) the density function of the random variable $\max \left\{ T_{1},T_{2}\right\} $ where, as in Eq. (\ref{311}), $T_{1}$ and $T_{2}$ are IID copies of the input $T$. With this density function at hand, the following formula is presented: 
\begin{equation}
\int_{0}^{\infty }\left[ \frac{M\left( \tau \right) -\mu }{\mu }\right] f_{\max }\left( \tau \right) d\tau =1-2\mathcal{I}_{Gini}.
\label{313}
\end{equation}
The derivation of Eq. (\ref{313}) is detailed in the Methods.

Eq. (\ref{313}) manifests the effect of the input's Gini index $\mathcal{I} _{Gini}$ on the output's mean $M\left( \tau \right) $. Indeed, Eq. (\ref{313}) yields the following pair of \emph{Gini-index criteria}:

\begin{enumerate}
\item[$\bullet $] If the input's Gini index is smaller than half, $\mathcal{I}_{Gini}<\frac{1}{2}$, then there exist timers $\tau $ for which sharp restart is detrimental.

\item[$\bullet $] If the input's Gini index is larger than half, $\mathcal{I}_{Gini}>\frac{1}{2}$, then there exist timers $\tau $ for which sharp restart is beneficial.
\end{enumerate}

These Gini-index criteria have a threshold form that is identical to that of the CV-index criteria of subsection \ref{22}. Specifically, the input's Gini index $\mathcal{I}_{Gini}$ -- the input's `Gini inequality' -- is compared to the threshold level $\frac{1}{2}$. To determine the existence of timers for which the application of the sharp-restart algorithm is detrimental or beneficial, one needs to check if the input's `Gini inequality' $\mathcal{I}_{Gini}$ is below or above the threshold level $\frac{1}{2}$, respectively.

On the one hand, $\mathcal{I}_{CV}$ and $\mathcal{I}_{Gini}$ are markedly different inequality indices of the input  $T$. On the other hand, these very different inequality indices yield very similar existence results regarding timers $\tau $ for which sharp restart is detrimental or beneficial. We shall elaborate on the relation between the CV-index criteria and the Gini-index criteria in the discussion at the end of this section.

\subsection{\label{32}Bonferroni-index criteria}

Not as popular as the Gini index, yet no less profound, is the \emph{Bonferroni index} \cite{Bon}-\cite{EG}. This subsection addresses the Bonferroni index $\mathcal{I}_{Bonf}$ of the input $T$.

In terms of the input's Lorenz curve, the input's Bonferroni index $\mathcal{I}_{Bonf}$ is the average of the normalized vertical distances $\left[ q-L\left( q\right) \right] /q$ ($0<q<1$) between the Lorenz curve and the diagonal line \cite{EG}. Namely,
\begin{equation}
\mathcal{I}_{Bonf}=\int_{0}^{1}\frac{q-L\left( q\right) }{q}dq.
\label{612}
\end{equation}
As noted in subsection \ref{23}, for any fixed number $q$, the normalized vertical distance $\left[ q-L\left( q\right) \right] /q$ is an inequality index. Hence, the Bonferroni index $\mathcal{I}_{Bonf}$ is an average of inequality indices.

The Bonferroni-index representation of \eref{612} exhibits no clear connection to Eq. (\ref{211}), the output's mean. To make the connection more apparent we use $\phi \left( \tau \right) =\mathbf{E}\left[T|T\leq \tau \right] $ -- the input's conditional mean, given the information that the input is no larger than the timer. In terms of the input's density and distribution functions, this conditional mean is given by: 
\begin{equation}
\phi \left( \tau \right) =\int_{0}^{\tau }t\frac{f\left( t\right) }{F\left(\tau \right) }dt=\frac{1}{F\left( \tau \right) }\int_{0}^{\tau }tf\left(t\right) dt.  \label{321}
\end{equation}
Evidently, this conditional mean is no larger than the input's mean: $\phi\left( \tau \right) \leq \mu $.

In terms of the conditional mean $\phi \left( \tau \right) $, the input's Bonferroni index $\mathcal{I}_{Bonf}$ admits the following representation \cite{EG}:
\begin{equation}
\mathcal{I}_{Bonf}=1-\frac{1}{\mu }\int_{0}^{\infty }\phi \left( \tau\right) f\left( \tau \right) d\tau.  \label{322}
\end{equation}
The integral appearing on the right-hand side of Eq. (\ref{322}) is a weighted average of the conditional mean $\phi \left( \tau \right) $, where the averaging is with respect to the input's density function. The Bonferroni-index representation of Eq. (\ref{322}) is based on the disparity between two terms: the weighted average of the conditional mean $\int_{0}^{\infty }\phi \left( \tau \right) f\left( \tau \right) d\tau$ versus the input's mean $\mu$. The smaller the disparity between the two terms -- the closer is the Bonferroni index to its zero lower bound. Conversely, the larger the disparity between the two terms -- the closer is the Bonferroni index to its unit upper bound.

In terms of the conditional mean $\phi \left( \tau \right) $, the output's mean of Eq. (\ref{211}) admits the following representation:
\begin{equation}
M\left( \tau \right) =\phi \left( \tau \right) +\tau \frac{\bar{F}\left(\tau \right) }{F\left( \tau \right) }. \label{323}
\end{equation}
The derivation of Eq. (\ref{323}) is detailed in the Methods. Both Eqs. (\ref{322}) and (\ref{323}) incorporate the term $\phi \left(\tau \right) $. This commonality suggests that a connection between the input's Bonferroni index $\mathcal{I}_{Bonf}$ and the output's mean $M\left( \tau \right) $ may exist. We shall now establish such a connection.

Introduce the value
\begin{equation}
\nu =\int_{0}^{\infty }t\frac{f\left( t\right) }{F\left( t\right) } dt=\int_{0}^{\infty }\ln \left[\frac{1}{F\left( t\right) }\right] dt.  \label{325}
\end{equation}
With the value $\nu$ at hand, the following formula is presented: 
\begin{equation}
\int_{0}^{\infty }\left[ \frac{M\left( \tau \right) -\mu }{\mu }\right] f\left( \tau \right) d\tau =\frac{\nu -\mu }{\mu }-\mathcal{I}_{Bonf},  \label{324}
\end{equation}
The derivation of Eq. (\ref{324}) is detailed in the Methods.

Eq. (\ref{324}) manifests the effect of the input's Bonferroni index $\mathcal{I}_{Bonf}$ on the output's mean $M\left( \tau \right) $. Indeed, setting the threshold level $l_{Bonf}=\frac{\nu -\mu }{\mu }$, Eq. (\ref{324}) yields the following pair of \emph{Bonferroni-index criteria}:

\begin{enumerate}
\item[$\bullet $] If the input's Bonferroni index is smaller than its threshold level, $\mathcal{I}_{Bonf}<l_{Bonf}$, then there exist timers $\tau $ for which sharp restart is detrimental.

\item[$\bullet $] If the input's Bonferroni index is larger than its threshold level, $\mathcal{I}_{Bonf}>l_{Bonf}$, then there exist timers $\tau $ for which sharp restart is beneficial.
\end{enumerate}

\noindent The Bonferroni-index criteria have a threshold form that is similar to the threshold form of the CV-index criteria of subsection \ref{22}, and to the threshold form of the Gini-index criteria of subsection \ref{31}. Specifically, the input's Bonferroni index $\mathcal{I}_{Bonf}$ -- the input's `Bonferroni inequality' -- is compared to the threshold level $l_{Bonf}=\frac{\nu -\mu }{\mu }$. To determine the existence of timers for which the application of the sharp-restart algorithm is detrimental or beneficial, one needs to check if the input's `Bonferroni inequality' $\mathcal{I}_{Bonf}$ is below or above the threshold level $l_{Bonf}$, respectively.

\subsection{\label{33} Discussion}

We conclude this section with remarks regarding the resemblance and the interplay between the CV criteria of subsection \ref{21}, and the Gini-index criteria of subsection \ref{31}. The remarks are based on the deviation $T_{1}-T_{2}$ between two IID copies, $T_{1}$ and $T_{2}$, of the input. 

It is straightforward to observe that the mean square deviation (MSD) between the two IID copies is twice the input's variance: $\mathbf{E} [\left\vert T_{1}-T_{2} \right\vert ^{2}]=2\sigma ^{2}$. Consequently, in terms of the MSD, the input's squared CV admits the following representation:
\begin{equation}
\frac{\sigma^{2}}{\mu^{2}} =\frac{\mathbf{E} [\left\vert T_{1}-T_{2} \right\vert ^{2}]} {2 \mu^{2}}.  \label{331}
\end{equation}

An alternative to the aforementioned MSD is the mean absolute deviation (MAD) between the two IID copies, $\mathbf{E} [\left\vert T_{1}-T_{2} \right\vert]$ \cite{YL}. From a planar-geometry perspective, the MSD manifests aerial distance, and the MAD manifests walking distance \cite{HarInq}. In terms of the MAD, the input's Gini index admits the following representation \cite{TouInq}:
\begin{equation}
\mathcal{I}_{Gini}=\frac{\mathbf{E} [\left\vert T_{1}-T_{2} \right\vert]}{2 \mu }.  \label{332}
\end{equation}

Eqs. (\ref{331}) and (\ref{332}) highlight the resemblance between the input's squared CV, $\sigma^{2} / \mu^{2}$, and the input's Gini index, $\mathcal{I}_{Gini}$. Jensen's inequality implies that the squared MAD is no larger than the MSD, $\mathbf{E} [\left\vert T_{1}-T_{2} \right\vert]^{2} \leq \mathbf{E} [\left\vert T_{1}-T_{2} \right\vert ^{2}]$. Consequently, Eqs. (\ref{331}) and (\ref{332}) imply the following relation between the input's Gini index and the input's CV: 
\begin{equation}
\sqrt{2} \cdot \mathcal{I}_{Gini} \leq \frac{\sigma}{\mu}. \label{333}
\end{equation}

The interplay between the CV criteria and the Gini-index criteria comprises four different scenarios (see Fig. 2): ``DD'', ``BB'', ``DB'' and ``BD''. In the ``DD'' and ``BB'' scenarios the Gini and CV criteria are in accord -- both asserting the existence of timers for which sharp restart is either detrimental (``DD'') or beneficial (``BB''). In the ``DB'' and ``BD'' scenarios the Gini and CV criteria are in dis-accord -- one criterion asserts the existence of timers for which sharp restart is detrimental, and another criterion asserts the existence of timers for which sharp restart is beneficial. The ``DB'' and ``BD'' scenarios underscore the fact that sharp restart can be detrimental for some timers, and beneficial for other timers \cite{MP1SR}. An illustrative example that demonstrates the ``DB'' scenario is described in Fig. 3.

\begin{figure}[t!]
\centering
\includegraphics[width=9cm]{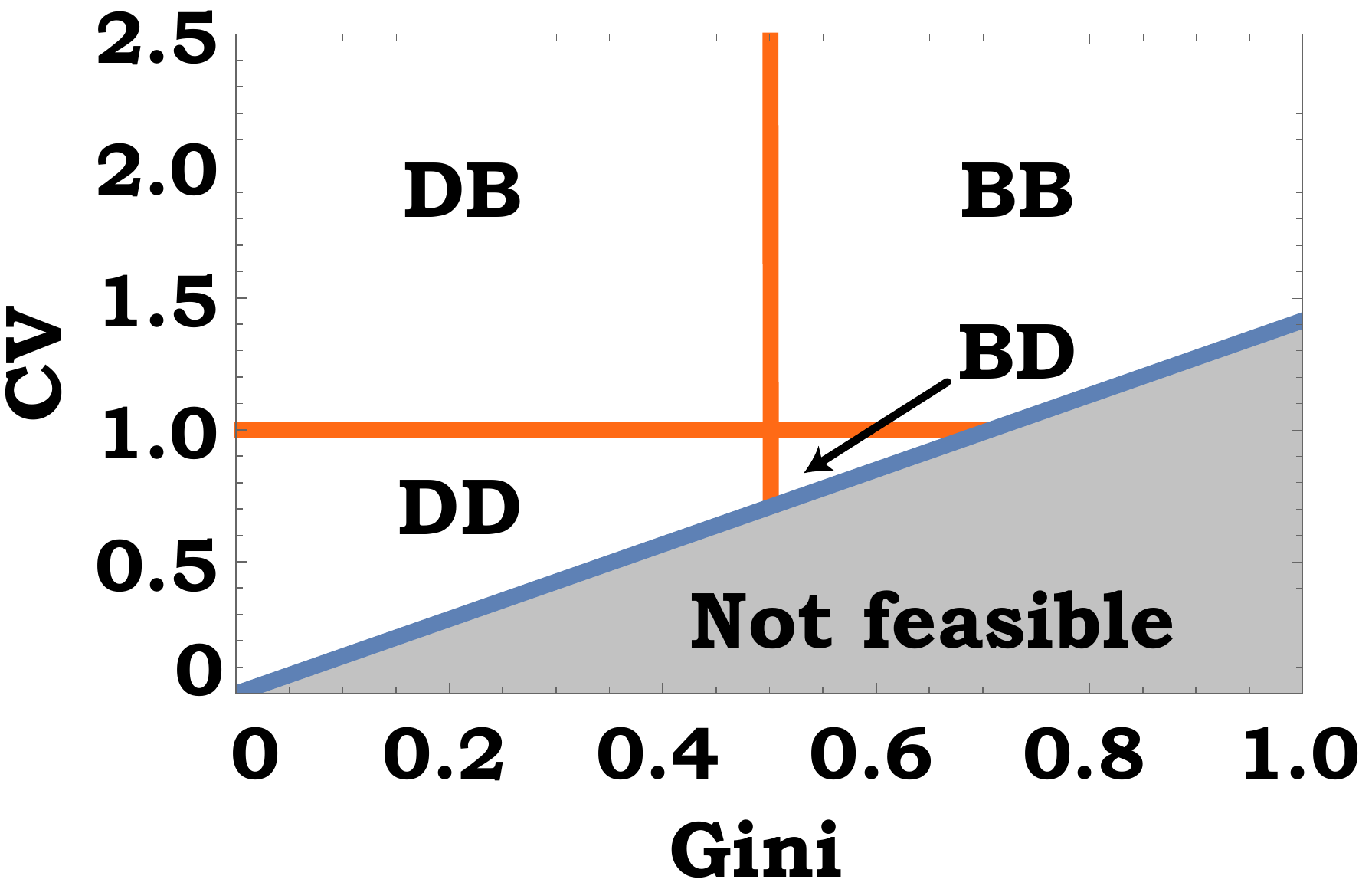}
\caption{A Gini-CV `phase diagram'. The horizontal axis represents the value of the input's Gini index $\mathcal{I}_{Gini}$, and the vertical axis represents the value of the input's CV $\frac{\sigma}{\mu}$. The line $\sqrt{2} \mathcal{I}_{Gini}= \frac{\sigma}{\mu}$ is depicted in blue; according to Eq. (\ref{333}), feasible pairs of Gini-CV values $(\mathcal{I}_{Gini},\frac{\sigma}{\mu})$ do not reside below this line. The Gini-CV `phase diagram' comprises four different regions -- each manifesting a different scenario. The trapezoid region ``DD'' ($0 \leq \mathcal{I}_{Gini}<\frac{1}{2}$ and $\sqrt{2}\mathcal{I}_{Gini} \leq \frac{\sigma}{\mu} <1 $): in this region both the CV criteria and the Gini-index criteria assert the existence of timers for which sharp restart is detrimental. The trapezoid region ``BB'' ( $ \frac{1}{2} < \mathcal{I}_{Gini} \leq 1 $ and $\sqrt{2}\mathcal{I}_{Gini} \leq \frac{\sigma}{\mu} < \infty$): in this region both the CV criteria and the Gini-index criteria assert the existence of timers for which sharp restart is beneficial. The rectangular region ``DB'' ($0 \leq \mathcal{I}_{Gini}<\frac{1}{2}$ and $1< \frac{\sigma}{\mu} < \infty$): in this region a Gini-index criterion asserts the existence of timers for which sharp restart is detrimental, and a CV criterion asserts the existence of timers for which sharp restart is beneficial. The triangular region ``BD'' ($ \frac{1}{2} < \mathcal{I}_{Gini} < \frac{1}{\sqrt2} $ and $\sqrt{2}\mathcal{I}_{Gini} \leq \frac{\sigma}{\mu} < 1$): in this region a Gini-index criterion asserts the existence of timers for which sharp restart is beneficial, and a CV criterion asserts the existence of timers for which sharp restart is detrimental.}
\label{Pareto}
\end{figure}

\begin{figure}[t!]
\centering
\includegraphics[width=9cm]{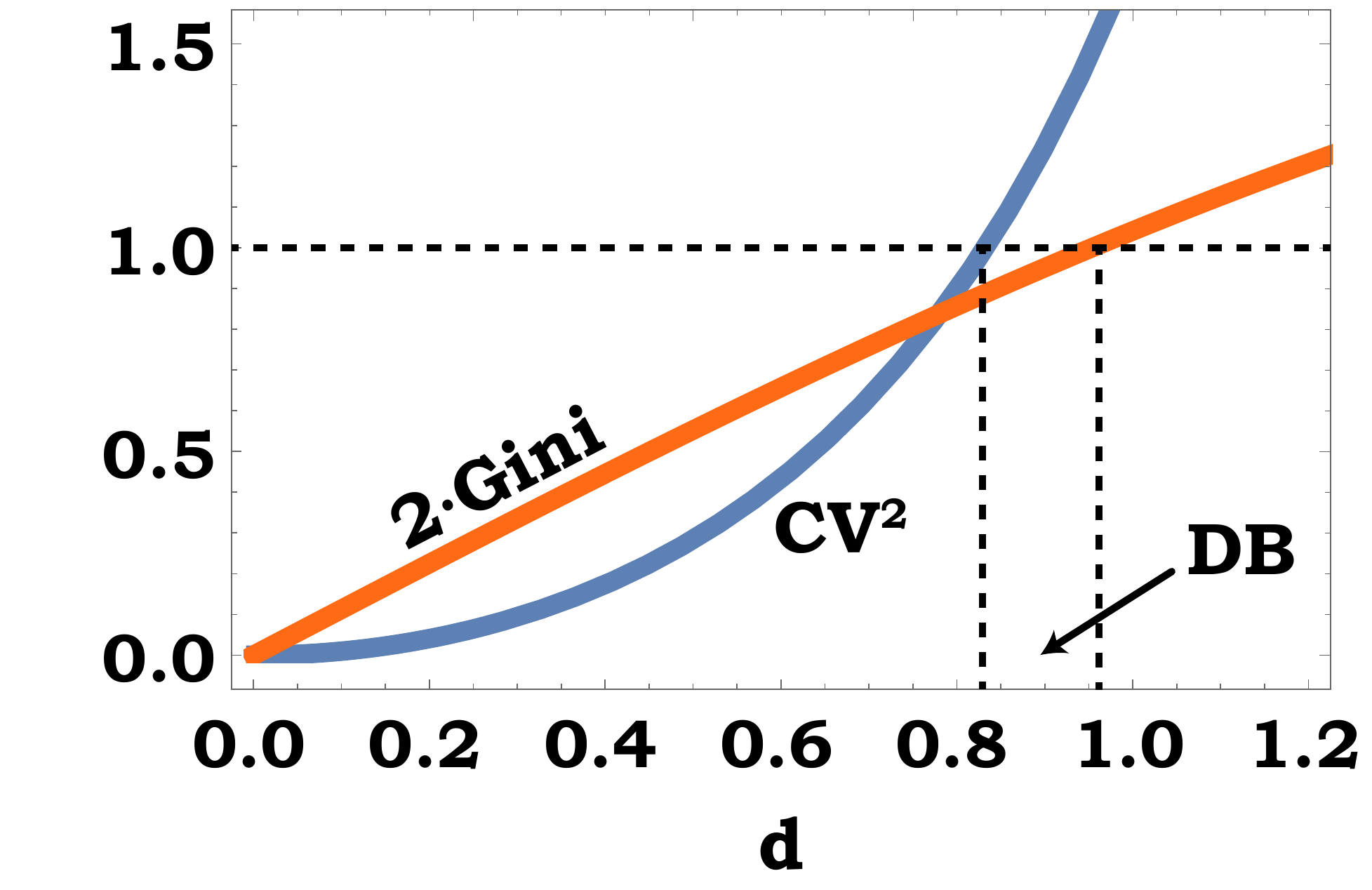}
\caption{Lognormal example of the ``DB'' scenario. Lognormal random variables are of major importance across the sciences \cite{Gal}-\cite{Dow}. In particular, Lognormal service times are widely observed in call centers \cite{BGM}-\cite{GT2}. In this example we consider the input $T$ to be Lognormal: $\ln{(T)}$ is a normal random variable with an arbitrary mean, and with standard deviation $d$. The squared CV of this input is $\exp{(d^2)}-1$, and the Gini index of this input is $\erf(d/2)$ \cite{LFS} (here $\erf(z)$ denotes the Gauss error function). As functions of the positive standard-deviation parameter $d$: the squared CV is depicted in blue, and twice the Gini index is depicted in orange. The CV is larger than one if and only if $d> \sqrt{\ln{(2)}}$, and the Gini index is smaller than half if and only if $d<2 \erf^{-1}(\frac{1}{2})$. Hence, the scenario ``DB'' holds when the standard-deviation parameter is in the range $\sqrt{\ln{(2)}}<d<2\erf^{-1}(\frac{1}{2})$, which is indicated in the figure.}
\label{Lognormal_fig}
\end{figure}

\section{\label{4}Mean and median timers}

The inequality-indices criteria established so far, in sections \ref{2} and \ref{3}, were existence results. Namely, these criteria determined the existence of timers $\tau $ for which the application of sharp restart is detrimental or beneficial. However, these criteria do not address particular timers $\tau $. Indeed, these criteria are incapable of determining if sharp restart with a particular timer $\tau $ is detrimental or beneficial. This section begins to address -- via socioeconomic inequality indices -- particular timers $\tau $. Specifically, this section shall address sharp restart with two specific and `natural' timers: a timer whose value is the input's mean, and a timer whose value is the input's median.

Along this section the following representation of the output's mean of Eq. (\ref{211}) shall be used:
\begin{equation}
M\left( \tau \right) =\frac{\mu +\tau -\mathbf{E}\left[ \left\vert T-\tau \right\vert \right] }{2F\left( \tau \right) }.  \label{400}
\end{equation}
The derivation of Eq. (\ref{400}) is detailed in the Methods. The term $\mathbf{E}\left[ \left\vert T-\tau \right\vert \right] $ appearing in Eq. (\ref{400}) is the mean absolute deviation (MAD) \cite{YL} between the input $T$ and the timer $\tau $.

\subsection{\label{41}Mean timer}

In this subsection we consider sharp restart with the particular timer $\tau=\mu$, the input's mean. Namely, a timer that equals the mean of the task's completion time $T$. As shall be shown below, sharp restart with this particular timer is intimately related to an inequality index of the input that is termed \emph{Pietra index} \cite{Pie}-\cite{SJ}.

From a functional-analysis perspective, \eref{611} implies that the input's Gini index $\mathcal{I}_{Gini}$ is based on the ``$L_{1}$ distance'' between the Lorenz curve $y=L(x)$ and the diagonal line $y=x$. Switching from the $L_{1}$ distance to the ``$L_{\infty}$ distance'' yields the input's Pietra index \cite{ES}: $\mathcal{I}_{Pietra}$ is the maximal vertical distance between the input's Lorenz curve $y=L\left( x\right) $ and the diagonal line $y=x$. Namely,
\begin{equation}
\mathcal{I}_{Pietra}=\max_{0\leq q\leq 1}\left[ q-L\left( q\right) \right].  \label{613}
\end{equation}

The Pietra-index representation of \eref{613} exhibits no clear connection to Eq. (\ref{211}), the output's mean. To make the connection more apparent we use the following, alternative, Pietra-index representation \cite{TouInq}:
\begin{equation}
\mathcal{I}_{Pietra}=\frac{1}{2\mu }\mathbf{E}\left[ \left\vert T-\mu\right\vert \right].  \label{411}
\end{equation}%
The Pietra-index representation of Eq. (\ref{411}) is based on the term $\mathbf{E}\left[ \left\vert T-\mu \right\vert \right] $, the MAD between the input $T$ and its mean $\mu $ \cite{Edd}. The smaller this MAD -- the closer is the Pietra index to its zero lower bound. Conversely, the larger this MAD -- the closer is the Pietra index to its unit upper bound.

Setting $\tau =\mu $ in Eq. (\ref{400}), and using Eq. (\ref{411}), yields the formula
\begin{equation}
\frac{M\left( \mu \right) -\mu }{\mu }=\frac{\bar{F}\left( \mu \right) - \mathcal{I}_{Pietra}}{F\left( \mu \right) }. \label{412}
\end{equation}
Eq. (\ref{412}) manifests, for the particular timer $\tau =\mu $, the effect of the input's Pietra index $\mathcal{I}_{Pietra}$ on the output's mean $M\left( \mu \right) $. Indeed, set the threshold level $l_{Pietra}=\bar{F}\left( \mu \right)$, the probability that the input exceeds its mean. Then, Eq. (\ref{412}) yields the following pair of \emph{Pietra-index criteria}:
\begin{enumerate}
\item[$\bullet $] Sharp restart with timer $\tau =\mu $ is detrimental if and only if the input's Pietra index is smaller than its threshold level, $\mathcal{I}_{Pietra}<l_{Pietra}$.

\item[$\bullet $] Sharp restart with timer $\tau =\mu $ is beneficial if and only if the input's Pietra index is larger than its threshold level, $\mathcal{I}_{Pietra}>l_{Pietra}$.
\end{enumerate}

Jensen's inequality implies that the squared Pietra-index MAD is no larger than the input's variance, i.e. $\mathbf{E}\left[ \left\vert T-\mu \right\vert \right]^{2} \leq \mathbf{E}[\left\vert T-\mu \right\vert ^{2}] =\sigma ^{2} $. Consequently, twice the input's Pietra index is no larger than the input's CV, $2 \cdot \mathcal{I}_{Pietra} \leq \frac{\sigma}{\mu}$. Also, the input's Pietra index is no larger than the input's Gini index, $ \mathcal{I}_{Pietra} \leq \mathcal{I}_{Gini}$ \cite{SocG1}.

These Pietra-CV and Pietra-Gini relations -- combined, respectively, with the CV criteria of subsection \ref{21}, and with the Gini-index criteria of subsection \ref{31} -- yield the following \emph{Pietra-index corollary}: If the input's Pietra index is larger than half, $\mathcal{I}_{Pietra}>\frac{1}{2}$, then there exist timers $\tau $ for which sharp restart is beneficial. In case the input's median is no-larger than the input's mean, $m \leq \mu$, the Pietra-index criteria `upgrade' the Pietra-index corollary as follows: If the input's Pietra index is larger than half, $\mathcal{I}_{Pietra}>\frac{1}{2}$, then sharp restart with timer $\tau =\mu $ is beneficial.

Last, with regard to being larger than the level $\frac{1}{2}$, we note the following resemblances between the CV, Gini, and Pietra indices. CV index: $\mathcal{I}_{CV}>\frac{1}{2}$ is equivalent to $\mathbf{E}[\left\vert T-\mu \right\vert ^{2}]  >\mu^{2}$. Gini index: $\mathcal{I}_{Gini}>\frac{1}{2} $ is equivalent to $ \mathbf{E} [\left\vert T_{1}-T_{2} \right\vert]>\mu$. Pietra index: $\mathcal{I}_{Pietra}>\frac{1}{2} $ is equivalent to $ \mathbf{E} [\left\vert T-\mu \right\vert]>\mu$. 

\subsection{\label{42}Median timer}

In this subsection we consider sharp restart with the particular timer $\tau=m$, the input's median. Namely, a timer that equals the median of the task's completion time $T$.\footnote{Specifically, the median $m$ is the unique positive value at which the input's distribution function and survival function intersect: $F\left(m\right) =\frac{1}{2}=\bar{F}\left( m\right) $. As the input's density function is considered to be positive-valued over the positive half-line, the input's median is well defined indeed.} As shall be shown below, sharp restart with this particular timer is intimately related to an inequality index of the input that is termed \emph{vertical-diameter index} (Vdiam index) \cite{SocG1}-\cite{SocG2}.

The description of the Vdiam index uses the Lorenz curve $y=L(x)$ and its corresponding complementary Lorenz curve, $y=\bar{L}(x)$ ($0\leq x,y\leq 1$) \cite{TouInq}. The socioeconomic meaning of the complementary Lorenz curve is: the top (rich) $100x\%$ of the society members possess $100y\%$ of the society's overall wealth. The complementary Lorenz curve $y=\bar{L}(x)$ has the same properties as the Lorenz curve $y=L(x)$, albeit one: it is concave rather than convex. Consequently, in the unit square, the complementary Lorenz curve $y=\bar{L}(x)$ is bounded from below by the diagonal line $y=x$.

As described above, the input's Pietra index $\mathcal{I}_{Pietra}$ is the ``$L_{\infty}$ distance'' between the input's Lorenz curve $y=L\left( x\right) $ and the diagonal line $y=x$. Switching from the ``$L_{\infty}$ distance'' between the Lorenz curve and the diagonal line, to the ``$L_{\infty}$ distance'' between the Lorenz curve and the complementary Lorenz curve, yields the input's Vdiam index \cite{SocG1}-\cite{SocG2}: $\mathcal{I}_{Vdiam}$ is the maximal vertical distance between the input's Lorenz curve $y=L\left( x\right) $ and the complementary Lorenz curve $y=\bar{L}(x)$. It turns out that $\mathcal{I}_{Vdiam}$ is the normalized vertical distance -- along the vertical line $x=\frac{1}{2}$ -- between the input's Lorenz curve $y=L\left( x\right) $ and the diagonal line $y=x$ \cite{SocG1}-\cite{SocG2}. Namely,
\begin{equation}
\mathcal{I}_{Vdiam}=1-2L\left( \frac{1}{2}\right).  \label{615}
\end{equation}%

The Vdiam-index representation of \eref{615} exhibits no clear connection to Eq. (\ref{211}), the output's mean. To make the connection more apparent we use the following, alternative, Vdiam index representation \cite{TouInq}:
\begin{equation}
\mathcal{I}_{Vdiam}=\frac{1}{\mu }\mathbf{E}\left[ \left\vert T-m\right\vert \right] ,  \label{421}
\end{equation}
where $m$ is the input's median. The Vdiam-index representation of Eq. (\ref{421}) is based on the term $\mathbf{E}\left[ \left\vert T-m \right\vert \right] $, the MAD between the input $T$ and its median $m$. The smaller this MAD -- the closer is the Vdiam index to its zero lower bound. Conversely, the larger this MAD -- the closer is the Vdiam index to its unit upper bound.

Setting $\tau =m$ in Eq. (\ref{400}), and using Eq. (\ref{421}), yields the formula
\begin{equation}
\frac{M\left( m\right) -\mu }{\mu }=\frac{m}{\mu }-\mathcal{I}_{Vdiam}. \label{422}
\end{equation}
Eq. (\ref{422}) manifests, for the particular timer $\tau =m$, the effect of the input's Vdiam index $\mathcal{I}_{Vdiam}$ on the output's mean $M\left(m\right) $. Indeed, set the threshold level $l_{Vdiam}=\frac{m}{\mu }$, the ratio of the input's median to the input's mean. Then, Eq. (\ref{422}) yields the following pair of \emph{Vdiam-index criteria}:

\begin{enumerate}
\item[$\bullet $] Sharp restart with timer $\tau =m$ is detrimental if and only if the input's Vdiam index is smaller than its threshold level, $\mathcal{I}_{Vdiam}<l_{Vdiam}$.

\item[$\bullet $] Sharp restart with timer $\tau =m$ is beneficial if and only if the input's Vdiam index is larger than its threshold level, $\mathcal{I}_{Vdiam}>l_{Vdiam}$.
\end{enumerate}

\subsection{\label{43} Discussion}

We conclude this section with remarks regarding the input's mean and median, the CV criteria of subsection \ref{21}, and the Vdiam-index criteria of subsection \ref{42}. The remarks are based on an optimization-problem perspective.

The ``$L_{p}$ distance'' between the input $T$ and the timer $\tau $ is 
\begin{equation}
D_{p}\left( \tau \right) =\left\{ \mathbf{E}\left[ \left\vert T-\tau \right\vert ^{p}\right] \right\} ^{1/p}.  \label{430}
\end{equation}
The parameter $p$ of the $L_{p}$ distance takes values in the range $p\geq 1$, and its most notable values are $p=1$ and $p=2$. Specifically, the $L_{1}$ distance is the aforementioned mean absolute deviation (MAD) $\mathbf{E}[\left\vert T-\tau \right\vert ]$ between the input $T$ and the timer $\tau $. And, the $L_{2}$ distance is the square root of the mean square deviation (MSD) $\mathbf{E}[\left\vert T-\tau \right\vert ^{2}]$ between the input $T$ and the timer $\tau $.

The inputs's mean and median are intimately related, respectively, to the $L_{2}$ and $L_{1}$ distances. Indeed, consider the optimization problem $\min_{0<\tau <\infty }D_{p}\left( \tau \right) $. This optimization problem seeks the timer $\tau$ that is closest -- in the $L_{p}$ distance -- to the input $T$. For the $L_{2}$ distance the minimum is attained at the input's mean, and the minimum value is the input's standard deviation: $\arg \min_{0<\tau <\infty }D_{2}\left( \tau \right) =\mu $ and $\min_{0<\tau <\infty }D_{2}\left( \tau \right) =\sigma $. For the $L_{1}$ distance the minimum is attained at the input's median, and the minimum value is the MAD between the input and its median: $\arg \min_{0<\tau <\infty }D_{1}\left( \tau \right) =m$ and $\min_{0<\tau <\infty }D_{1}\left( \tau \right) =\mathbf{E}\left[ \left\vert T-m\right\vert \right]$.

The CV criteria of subsection \ref{21} assert that -- in order to determine the existence of timers $\tau $ for which the sharp restart algorithm is detrimental or beneficial -- one has to compare the input's standard deviation $\sigma $ to the input's mean $\mu $. In other words, the CV criteria compare the minimal value $\min_{0<\tau <\infty}D_{2}\left( \tau \right) $ to the minimizing point $\arg \min_{0<\tau <\infty
}D_{2}\left( \tau \right) $.

Eq. (\ref{422}) can be re-written as follows: 
\begin{equation}
M\left( m\right) -\mu =m-\mathbf{E}\left[ \left\vert T-m\right\vert \right]. \label{431}
\end{equation}
Consequently, the Vdiam-index criteria of subsection \ref{42} -- regarding restart with the particular timer $\tau =m$ -- can be re-formulated as follows: sharp restart is detrimental if and only if $\mathbf{E}\left[\left\vert T-m\right\vert \right] < m$, and is beneficial if and only if $\mathbf{E}\left[ \left\vert T-m\right\vert \right] > m $. In other words, the  Vdiam-index criteria compare the minimal value $ \min_{0<\tau <\infty }D_{1}\left( \tau \right) $ to the minimizing point $\arg \min_{0<\tau <\infty }D_{1}\left( \tau \right) $.

Thus, from the perspective of the optimization problem $\min_{0<\tau <\infty }D_{p}\left( \tau \right) $, there is an analogy between the CV criteria of subsection \ref{21} and the Vdiam-index criteria of subsection \ref{42}. Indeed, as pointed out above, both these criteria compare the minimal value $\min_{0<\tau <\infty}D_{p}\left( \tau \right) $ to the minimizing point $\arg \min_{0<\tau <\infty}D_{p}\left( \tau \right) $. Notably, the CV criteria and the Vdiam-index criteria are based on rather `neat' formulae -- Eq. (\ref{212}) and Eq. (\ref{431}), respectively.

\section{\label{5}General timers}

The previous section addressed two particular timers: $\tau =\mu $, where $\mu $ is the input's mean; and $\tau =m$, where $m$ is the input's median. To each of these timers a specific inequality index of the input was matched, and using these inequality indices it was determined when sharp restart (with these timers) is detrimental or beneficial. This section goes beyond the particular mean and median timers, and it addresses general timers $0<\tau <\infty $.

\subsection{\label{50}Sampling}

As noted in subsection \ref{22}, in order to measure the inherent `socioeconomic inequality' of the input $T$, this random variable was deemed to be the wealth of a randomly-sampled member of a virtual society. Now, with respect to this virtual society, consider also a randomly-sampled Dollar. Namely, sample at random a single Dollar from the society's overall wealth, and set $T_{dol}$ to be the wealth of the society member to whom the randomly-sampled Dollar belongs \cite{TouInq}. 

A temporal description of the random variable $T_{dol}$ is as follows. Perform, repeatedly and independently, $n$ rounds of the task under consideration (whose completion time is the random variable $T$). Specifically: starting at time $t=0$, perform the task for the first round; then, upon the first completion, start performing the task for the second round; then, upon the second completion, start performing the task for the third round; and continue so on and so forth for $n$ consecutive rounds. Denoting by $\left\{ T_{1},\cdots,T_{n} \right\} $ the durations of the tasks -- these durations being IID copies of the input $T$ -- we obtain that: $T_{1}$ is the completion time of the first round, $T_{1}+T_{2}$ is the completion time of the second round, ... , and $T_{1}+ \cdots +T_{n}$ is the completion time of the last round.

Now, place an observer at a random time epoch along the temporal interval $[0,T_{1}+ \cdots +T_{n}]$. This random placement of the observer implies that: the observer samples the task that is performed at round $k$ with probability $T_{k} / (T_{1}+ \cdots +T_{n}) $. In the limit $n\to\infty$, the duration of the task that is sampled by the observer converges, in law, to the random variable $T_{dol}$.

As the input $T$, also $T_{dol}$ is a positive-valued random variable. In terms of the input's density function and mean, the density function of the random variable $T_{dol}$ is $f_{dol}=\frac{1}{\mu}tf(t)$ ($t>0$) \cite{TouInq}. In turn, the distribution and survival functions of the random variable $T_{dol}$ are, respectively:
\begin{equation}
F_{dol}\left( t\right) =\frac{1}{\mu }\int_{0}^{t}sf\left( s\right) ds
\label{501}
\end{equation}
($t\geq 0$), and
\begin{equation}
\bar{F}_{dol}\left( t\right) =\frac{1}{\mu }\int_{t}^{\infty }sf\left(s\right) ds  \label{502}
\end{equation}
($t\geq 0$).

The input's Lorenz curve $y=L\left( x\right) $ couples together the input's distribution function $F\left( t\right) $, and the distribution function $F_{dol}\left( t\right) $ of the random variable $T_{dol}$. Indeed, the socioeconomic definitions of the Lorenz curve $y=L\left( x\right) $, and of the distribution functions $F\left( t\right) $ and $F_{dol}\left( t\right) $, implies that \cite{SocG1}-\cite{SocG2}:
\begin{equation}
L\left[ F\left( t\right) \right] =F_{dol}\left( t\right)  \label{600a}
\end{equation}
($t\geq 0$). An alternative formulation of Eq. (\ref{600a}) is:
\begin{equation}
F(t) =L^{-1}[F_{dol}(t)]  \label{600b}
\end{equation}
($t\geq 0$), where $x=L^{-1}(y)$ denotes the inverse function of the input's Lorenz curve $y=L\left( x\right) $. 

\subsection{\label{51}Vertical-index criteria}

Subsection \ref{23} introduced, for a fixed number $q$ (where $0<q<1$), the inequality index $[q-L\left(q\right) ]/q$: the `normalized' vertical distance -- along the vertical line $x=q$ -- between the Lorenz curve $y=L\left( x\right) $ and the diagonal line $y=x$. With respect to the input's distribution function $F\left( t\right) $, set $q$ to be the quantile corresponding to the timer $\tau $, i.e. $q=F\left( \tau \right) $. Eq. (\ref{600a}) implies that
\begin{equation}
\frac{q-L\left( q\right) }{q}=1-\frac{F_{dol}\left( \tau \right) }{F\left(\tau \right)}. \label{601}
\end{equation}
The quantity appearing in Eq. (\ref{601}) is an inequality index with an underpinning vertical Lorenz-curve geometric meaning. This quantity is henceforth termed the input's \emph{vertical index}, and is denoted $\mathcal{I}_{V}\left(\tau \right) $.  

Introduce the threshold level
\begin{equation}
l_{V}\left( \tau \right)= \frac{\tau }{\mu }\frac{\bar{F}\left( \tau \right) }{F\left( \tau \right) }.\label{6010}
\end{equation}
With the vertical index $\mathcal{I}_{V}\left(\tau \right) $ of Eq. (\ref{601}) at hand, as well as the threshold level $l_{V}\left( \tau \right)$ of Eq. (\ref{6010}), the following formula is presented:
\begin{equation}
\frac{M\left( \tau \right) -\mu }{\mu }=l_{V}\left( \tau \right)-\mathcal{I}_{V}\left(\tau \right).
\label{511}
\end{equation}
The derivation of Eq. (\ref{511}) is detailed in the Methods.

Eq. (\ref{511}) manifests the effect of the input's vertical index $\mathcal{I}_{V}\left( \tau \right) $ on the output's mean $M\left( \tau \right)$. Indeed, Eq. (\ref{511}) yields the following pair of \emph{vertical-index criteria}:

\begin{enumerate}
\item[$\bullet $] Sharp restart with timer $\tau $ is detrimental if and only if the input's vertical index is smaller than its threshold level, $\mathcal{I}_{V}\left( \tau \right) <l_{V}\left( \tau \right)$.

\item[$\bullet $] Sharp restart with timer $\tau $ is beneficial if and only if the input's vertical index is larger than its threshold level, $\mathcal{I}_{V}\left( \tau \right) >l_{V}\left( \tau \right)$.
\end{enumerate}

In particular, setting $\tau=\mu$ in the vertical-index criteria yields the Pietra-index criteria of subsection \ref{41}. And, setting $\tau=m$ in the vertical-index criteria yields the Vdiam-index criteria of subsection \ref{42}.

\subsection{\label{52}Horizontal-index criteria}

The vertical-index criteria of the previous subsection are based on the vertical distance between the Lorenz curve $y=L\left( x\right) $ and the diagonal line $y=x$. This subsection shifts from the vertical-distance perspective to a horizontal-distance perspective. 

For a fixed number $q$ (where $0<q<1$), consider the horizontal line $y=q$ of the unit square. The horizontal distance -- along the horizontal line $y=q$ -- between the Lorenz curve $y=L\left( x\right) $ and the diagonal line $y=x$ is: $L^{-1}\left( q\right) -q$.\footnote{As in Eq. (\ref{600b}) above, $x=L^{-1}(y)$ denotes the inverse function of the input's Lorenz curve $y=L\left( x\right) $.} This horizontal distance takes values in the range $\left[ 0,1-q\right] $. Consequently, the `normalized' horizontal distance $[L^{-1}\left( q\right)-q]/\left( 1-q\right) $ takes values in the unit interval $\left[ 0,1\right] $. It is straightforward to check that the normalized horizontal distance $[L^{-1}\left( q\right)-q]/\left( 1-q\right) $ meets the three inequality-index properties that were postulated in subsection \ref{22}. 

With respect to the distribution function $F_{dol}\left( t\right) $ of the random variable $T_{dol}$, set $q$ to be the quantile corresponding to the timer $\tau $, i.e. $q=F_{dol}\left( \tau \right) $. Eq. (\ref{600b}) implies that
\begin{equation}
\frac{L^{-1}\left( q\right) -q}{1-q}=1-\frac{\bar{F}\left( \tau \right) }{\bar{F}_{dol}\left( \tau \right) }.  \label{602}
\end{equation}
The quantity appearing in Eq. (\ref{602}) is an inequality index with an underpinning horizontal Lorenz-curve geometric meaning. This quantity is henceforth termed the input's \emph{horizontal index}, and is denoted $\mathcal{I}_{H}\left(\tau \right) $.  

Introduce the threshold level
\begin{equation}
l_{H}\left( \tau \right)= \frac{\tau}{\tau+\mu}.  \label{6020}
\end{equation}
With the horizontal index $\mathcal{I}_{H}\left(\tau \right)$ of Eq. (\ref{602}) at hand, as well as the threshold level $l_{H}\left( \tau \right)$ of Eq. (\ref{6020}), we present the following formula:
\begin{equation}
\frac{M\left( \tau \right) -\mu }{\mu } =  \frac{(\tau+\mu) \bar{F}_{dol}(\tau)}{\mu F(\tau)}  \cdot\left[ l_{H}\left( \tau \right)-\mathcal{I}_{H}\left(\tau \right) \right].  \label{521}
\end{equation}
The derivation of Eq. (\ref{521}) is detailed in the Methods. 

Eq. (\ref{521}) manifests the effect of the input's horizontal index $\mathcal{I}_{H}\left( \tau \right) $ on the output's mean $M\left( \tau \right) $. Indeed, Eq. (\ref{521}) yields the following pair of \emph{horizontal-index criteria}:

\begin{enumerate}
\item[$\bullet $] Sharp restart with timer $\tau $ is detrimental if and only if the input's horizontal index is smaller than its threshold level, $\mathcal{I}_{H}\left( \tau \right) <l_{H}\left( \tau \right) $.

\item[$\bullet $] Sharp restart with timer $\tau $ is beneficial if and only if the input's horizontal index is larger than its threshold level, $\mathcal{I}_{H}\left( \tau \right) >l_{H}\left( \tau \right) $.
\end{enumerate}

In particular, setting $\tau=\mu$ in the horizontal-index criteria yields the Pietra-index criteria of subsection \ref{41}. 

\subsection{\label{53}Discussion}

We conclude this section with remarks regarding the vertical-index criteria and the horizontal-index criteria. Also, using the horizontal-index criteria, we shall elaborate on a particular timer: $\tau=m_{dol}$, the median of the random variable $T_{dol}$.

In this section, given a positive timer $\tau $, two inequality indices of the input $T$ were matched to the timer: the vertical index $\mathcal{I}_{V}\left( \tau \right) $ of Eq. (\ref{601}), and the horizontal index $\mathcal{I}_{H}\left( \tau \right) $ of Eq. (\ref{602}). The vertical index $\mathcal{I}_{V}\left( \tau \right) $ quantifies the disparity, at the time point $t=\tau $, between the distribution functions of the random variables $T$ and $T_{dol}$. Similarly, the horizontal index $\mathcal{I} _{H}\left( \tau \right) $ quantifies the disparity, at the time point $t=\tau $, between the survival functions of the random variables $T$ and $T_{dol}$.

Accompanying the input's vertical and horizontal indices, $\mathcal{I}_{V}\left( \tau \right) $ and $\mathcal{I}_{H}\left( \tau \right) $, are corresponding threshold levels: $l_{V}\left( \tau \right) $ of Eq. (\ref{6010}), and $l_{H}\left( \tau \right) $ of Eq. (\ref{6020}). The comparisons between the vertical and horizontal indices and their corresponding threshold levels determines if sharp restart with the timer $\tau $ is detrimental or beneficial. While these comparisons are equivalent, they provide two different geometric Lorenz-curve perspectives -- vertical and horizontal.

The Pietra index of subsection \ref{41} emanated from maximizing the vertical distance between the Lorenz curve $y=L(x)$ and the diagonal line $y=x$. Analogously, the Vdiam index of subsection \ref{41} emanated from maximizing the vertical distance between the Lorenz curve $y=L(x)$ and the complementary Lorenz curve $y=\bar{L}(x)$. Shifting from the vertical perspective to the horizontal perspective has the following effects. As in the vertical perspective, maximizing the horizontal distance between the Lorenz curve $y=L(x)$ and the diagonal line $y=x$ yields the Pietra index \cite{SocG1}-\cite{SocG2}. However, maximizing the horizontal distance between the Lorenz curve $y=L(x)$ and the complementary Lorenz curve $y=\bar{L}(x)$ yields an inequality index of the input that is termed \emph{horizontal-diameter index} \cite{SocG1}-\cite{SocG2}.

It turns out that the input's Hdiam index $\mathcal{I}_{Hdiam}$ is the normalized horizontal distance -- along the horizontal line $y=\frac{1}{2}$ -- between the input's Lorenz curve $y=L\left( x\right) $ and the diagonal line $y=x$ \cite{SocG1}-\cite{SocG2}. Namely,
\begin{equation}
\mathcal{I}_{Hdiam}=2L^{-1}\left( \frac{1}{2}\right) -1.  \label{616}
\end{equation}
This index is intimately related to the median $m_{dol}$ of the random variable $T_{dol}$. Indeed, Eqs. (\ref{602}) and (\ref{616}) imply that $\mathcal{I}_{Hdiam}=\mathcal{I}_{H}\left(m_{dol} \right) $.

The Hdiam index has several representations. One of the input's Hdiam-index representations is \cite{TouInq}:
\begin{equation}
\mathcal{I}_{Hdiam}=\frac{1}{m_{dol}}\mathbf{E}\left[ \left\vert T-m_{dol}\right\vert \right].  \label{617}
\end{equation}%
The Hdiam-index representation of Eq. (\ref{617}) is based on the term $\mathbf{E}\left[ \left\vert T-m_{dol} \right\vert \right] $, the MAD between the input $T$ and the median $m_{dol}$ (which is the median of the random variable $T_{dol}$, not of the input $T$). The smaller this MAD -- the closer is the Vdiam index to its zero lower bound. Conversely, the larger this MAD -- the closer is the Vdiam index to its unit upper bound.

Setting $\tau =m_{dol}$ in Eq. (\ref{521}), and using Eq. (\ref{6020}), yields the formula
\begin{equation}
\frac{M\left( m_{dol}\right) -\mu }{\mu }=\frac{m_{dol}}{m_{dol} + \mu }-\mathcal{I}_{Hdiam}. \label{618}
\end{equation}
Eq. (\ref{618}) manifests, for the particular timer $\tau =m_{dol}$, the effect of the input's Hdiam index $\mathcal{I}_{Hdiam}$ on the output's mean $M\left(m_{dol}\right) $. Indeed, set the threshold level $l_{Hdiam}=\frac{m_{dol}}{m_{dol} + \mu }$. Then, Eq. (\ref{618}) yields the following pair of \emph{Hdiam-index criteria}:

\begin{enumerate}
\item[$\bullet $] Sharp restart with timer $\tau =m_{dol}$ is detrimental if and only if the input's Hdiam index is smaller than its threshold level, $\mathcal{I}_{Hdiam}<l_{Hdiam}$.

\item[$\bullet $] Sharp restart with timer $\tau =m_{dol}$ is beneficial if and only if the input's Hdiam index is larger than its threshold level, $\mathcal{I}_{Hdiam}>l_{Hdiam}$.
\end{enumerate}

\section{\label{7} Conclusion}

Following up on \cite{MP1SR}, in this paper we continued exploring the effect of sharp restart on mean performance. Using a positive deterministic timer $\tau$, sharp restart is applied to a general task whose completion time is a positive random variable $T$. Namely, initiating the task at time $t=0$, the task is restarted -- as long as it is not accomplished -- at the fixed time epochs $t=\tau,2\tau,3\tau,\cdots$. The mean completion time of the task `under restart', $M\left( \tau \right)$, is compared to the task's mean completion time, $\mathbf{E}\left[ T \right]=\mu$. This comparison determines if sharp restart improves mean performance, $M\left( \tau \right)<\mu$, or if it worsens mean performance, $M\left( \tau \right)>\mu$.

The analysis presented in this paper established that inequality indices of the random variable $T$ hold a treasure trove of information regarding the effect of sharp restart on mean performance. We showed that three inequality indices -- CV, Gini, and Bonferroni -- determine the very existence of timers with which sharp restart improves/worsens mean performance. Given a specific timer $\tau$, we further showed that there are inequality indices that relate to this timer, and that these inequality indices determine if sharp restart with the specific timer $\tau$ improves/worsens mean performance.

The novel results established in this paper provide a detailed `inequality roadmap' for sharp restart: eight pairs of universal inequality criteria that determine the effect of sharp restart on mean performance. The underpinning inequality indices are summarized in Table 1, and the criteria are summarized in Table 2. Each pair of criteria comprises a specific inequality index $\mathcal{I}$, and a corresponding threshold level $l$. All eight pairs of criteria share in common the following threshold pattern. If the index is larger than its corresponding threshold level then mean performance improves: $\mathcal{I}>l \Rightarrow M\left(\tau \right) <\mu$. And, if the index is smaller than its corresponding threshold level then mean performance worsens: $\mathcal{I}<l \Rightarrow M\left(\tau \right) >\mu$. 

On the one hand, inequality indices of the random variable $T$ measure the statistical heterogeneity of the task's completion time. On the other hand, it is known that sharp restart can match the mean-performance of any other restart protocol \cite{FPUR1,FPUR2}. Hence, combining these two facts together, we obtain the following take-home-message: restart improves mean performance when the underlying statistical heterogeneity is high; and it worsens mean performance when the underlying statistical heterogeneity is low. The universal inequality criteria established here articulate the take-home-message with unprecedented mathematical precision and resolution.

The first two pairs of inequality criteria -- CV and Gini -- have a fixed threshold level ($l=\frac{1}{2}$) that is independent of the task's completion time. Conversely, the latter six pairs of inequality criteria have threshold levels that depend on the task's completion time. For the latter six pairs of inequality criteria, the common threshold pattern admits an alternative interpretation that is described as follows. Given the value of the inequality index $\mathcal{I}$, there is a critical `mean value' $\mu_{c}$ that is based on $\mathcal{I}$. If the task's mean completion time is larger than the critical value then mean performance improves: $\mu > \mu_{c}\Rightarrow M\left(\tau \right) <\mu$. And, if the task's mean completion time is smaller than the critical value then mean performance worsens: $\mu < \mu_{c} \Rightarrow M\left(\tau \right) >\mu$. The critical values $\mu_{c}$ are specified in the right column of Table 2.

The inequality indices that underpin the inequality criteria have various equivalent representations \cite{TouInq}. As specified in Table 1, one such representation is based on mean square/absolute deviations (MSD/MAD) of the random variable $T$, the task's completion time; these deviations can be easily and reliably estimated from data. Thus, the application of the inequality criteria whose underpinning inequality indices admit a MSD/MAD representation is very practical, and can be done even when the distribution of the task's completion time is not known in full detail. Indeed, for example: scientists from various disciplines are well accustomed to estimating the CV, and estimating the Gini index is common practice in economics and in the social sciences; consequently, the use of the CV and Gini criteria should be straightforward to all.

Given a task, a `natural' sharp-restart timer to consider is the mean of the task's completion time, $\tau=\mu$. The Pietra criteria relate this natural timer, determining if sharp restart with this timer is beneficial or detrimental. In turn, building on relations between the CV, Gini, and Pietra inequality indices, the two following universal Pietra corollaries were also established. If the Pietra index is larger than the level $\frac{1}{2}$ then there exist timers with which sharp restart is beneficial. Moreover, if -- in addition to the Pietra index being larger than the level $\frac{1}{2}$ -- the input's median is no-larger than the input's mean, $m \leq \mu$, then: sharp restart with the specific timer $\tau =\mu$ is beneficial. As the Pietra index admits a MAD representation, the Pietra criteria and corollaries are very practical and are straightforward to use.

This paper unveiled profound connections between two seemingly unrelated topics: the measurement of socioeconomic inequality on the one hand, and the mean performance of sharp restart on the other hand. These connections provide researchers -- theoreticians and practitioners alike -- a whole new inequality vantage point, as well as a whole new inequality roadmap and toolbox, to work with in the interdisciplinary field of restart.\\

\newpage

\begin{center}
\bigskip {\Large Table 1}
\end{center}

\begin{center}
\begin{tabular}{|c|c|c|}
\hline
$%
\begin{array}{c}
\text{ \textbf{Inequality}} \\ 
\text{\textbf{index}}%
\end{array}%
$ & $%
\begin{array}{c}
\text{ \textbf{MSD/MAD}} \\ 
\text{\textbf{representation}}%
\end{array}%
$ & $%
\begin{array}{c}
\text{ \textbf{Lorenz}} \\ 
\text{\textbf{representation}}%
\end{array}%
$ \\ \hline
$%
\begin{array}{c}
\text{ } \\ 
\text{CV } \frac{\mathcal{I}_{CV}}{1-\mathcal{I}_{CV}}= \\ 
\text{ }%
\end{array}%
$ & $ \frac{1}{2\mu ^{2}}\mathbf{E}\left[ \left\vert
T_{1}-T_{2}\right\vert ^{2}\right] $ & --- \\ \hline
$%
\begin{array}{c}
\text{ } \\ 
\text{Gini }\mathcal{I}_{Gini}= \\ 
\text{ }%
\end{array}%
$ & $\frac{1}{2\mu }\mathbf{E}\left[ \left\vert T_{1}-T_{2}\right\vert %
\right] $ & $2\int_{0}^{1}\left[ q-L\left( q\right) \right] dq$ \\ \hline
$%
\begin{array}{c}
\text{ } \\ 
\text{Bonferroni }\mathcal{I}_{Bonf}= \\ 
\text{ }%
\end{array}%
$ & --- & $\int_{0}^{1}\frac{q-L\left( q\right) }{q}dq$ \\ \hline
$%
\begin{array}{c}
\text{ } \\ 
\text{Pietra }\mathcal{I}_{Pietra}= \\ 
\text{ }%
\end{array}%
$ & $\frac{1}{2\mu }\mathbf{E}\left[ \left\vert T-\mu \right\vert \right] $
& $\max_{0\leq q\leq 1}\left[ q-L\left( q\right) \right] $ \\ \hline
$%
\begin{array}{c}
\text{ } \\ 
\text{Vdiam }\mathcal{I}_{Vdiam}= \\ 
\text{ }%
\end{array}%
$ & $\frac{1}{\mu }\mathbf{E}\left[ \left\vert T-m\right\vert \right] $ & $%
1-2L\left( \frac{1}{2}\right) $ \\ \hline
$%
\begin{array}{c}
\text{ } \\ 
\text{Hdiam }\mathcal{I}_{Hdiam}= \\ 
\text{ }%
\end{array}%
$ & $\frac{1}{m_{dol}}\mathbf{E}\left[ \left\vert T-m_{dol}\right\vert %
\right] $ & $2L^{-1}\left( \frac{1}{2}\right) -1$ \\ \hline
$%
\begin{array}{c}
\text{ } \\ 
\text{Vertical }\mathcal{I}_{V}\left( \tau \right)=  \\ 
\text{ }%
\end{array}%
$ & --- & $\frac{q-L\left( q\right) }{q}$ \\ \hline
$%
\begin{array}{c}
\text{ } \\ 
\text{Horizontal }\mathcal{I}_{H}\left( \tau \right)=  \\ 
\text{ }%
\end{array}%
$ & --- & $\frac{L^{-1}\left( q\right) -q}{1-q}$ \\ \hline
\end{tabular}

\bigskip 
\end{center}

\textbf{Table 1}: The eight inequality indices that underpin, respectively,
the eight pairs of universal inequality criteria for sharp restart. These inequality indices have various equivalent representations \cite{TouInq}, two of which are presented in the table: representations based on mean square/absolute deviations (MSD/MAD) of the task's completion time $T$; and representations based on the Lorenz curve $L(\cdot)$ of the task's completion time $T$ (see section \ref{23} for the details).
Remarks regarding specific inequality indices are the following. CV and
Gini: $T_{1}$ and $T_{2}$ are two IID copies of the random variable $T$. Vdiam: $m$ is the median of the random
variable $T$. Hdiam: $m_{dol}$ is the median of the random variable $T_{dol}$
(see section \ref{50} for the details). Vertical: $q=F\left( \tau \right) $,
where $F\left( \cdot \right) $ is the distribution function of the random
variable $T$. Horizontal: $q=F_{dol}\left( \tau \right) $, where $%
F_{dol}\left( \cdot \right) $ is the distribution function of the random
variable $T_{dol}$ (see section \ref{50} for the details). 

\newpage

\begin{center}
{\Large Table 2}
\end{center}

\begin{center}
\begin{tabular}{|c|c|c|c|}
\hline
$%
\begin{array}{c}
\text{\textbf{Inequality} } \\ 
\text{\textbf{index }}\mathcal{I=}%
\end{array}%
$ & $%
\begin{array}{c}
\text{ \textbf{Threshold}} \\ 
\text{\textbf{level }}l=%
\end{array}%
$ & $%
\begin{array}{c}
\text{ } \\ 
\text{\textbf{Timer}} \\ 
\text{ }%
\end{array}%
$ & $%
\begin{array}{c}
\text{ \textbf{Critical}} \\ 
\text{\textbf{value }}\mu _{c}=%
\end{array}%
$ \\ \hline
$%
\begin{array}{c}
\text{ } \\ 
\mathcal{I}_{CV} \\ 
\text{ }%
\end{array}%
$ & $\frac{1}{2}$ & Existence & ----- \\ \hline
$%
\begin{array}{c}
\text{ } \\ 
\mathcal{I}_{Gini} \\ 
\text{ }%
\end{array}%
$ & $\frac{1}{2}$ & Existence & ----- \\ \hline
$%
\begin{array}{c}
\text{ } \\ 
\mathcal{I}_{Bonf} \\ 
\text{ }%
\end{array}%
$ & $\frac{\nu -\mu }{\mu }$ & Existence & $\frac{\nu }{1+\mathcal{I}_{Bonf}}
$ \\ \hline
$%
\begin{array}{c}
\text{ } \\ 
\mathcal{I}_{Pietra} \\ 
\text{ }%
\end{array}%
$ & $\bar{F}\left( \mu \right) $ & $\tau =\mu $ & $\bar{F}^{-1}\left( 
\mathcal{I}_{Pietra}\right) $ \\ \hline
$%
\begin{array}{c}
\text{ } \\ 
\mathcal{I}_{Vdiam} \\ 
\text{ }%
\end{array}%
$ & $\frac{m}{\mu }$ & $\tau =m$ & $\frac{m}{\mathcal{I}_{Vdiam}}$ \\ \hline
$%
\begin{array}{c}
\text{ } \\ 
\mathcal{I}_{Hdiam} \\ 
\text{ }%
\end{array}%
$ & $\frac{m_{dol}}{m_{dol}+\mu }$ & $\tau =m_{dol}$ & $m_{dol}\frac{1-%
\mathcal{I}_{Hdiam}}{\mathcal{I}_{Hdiam}}$ \\ \hline
$%
\begin{array}{c}
\text{ } \\ 
\mathcal{I}_{V}\left( \tau \right)  \\ 
\text{ }%
\end{array}%
$ & $\frac{\tau }{\mu }\frac{\bar{F}\left( \tau \right) }{F\left( \tau
\right) }$ & $0<\tau <\infty $ & $\frac{\tau }{\mathcal{I}_{V}\left( \tau
\right) }\frac{\bar{F}\left( \tau \right) }{F\left( \tau \right) }$ \\ \hline
$%
\begin{array}{c}
\text{ } \\ 
\mathcal{I}_{H}\left( \tau \right)  \\ 
\text{ }%
\end{array}%
$ & $\frac{\tau }{\tau +\mu }$ & $0<\tau <\infty $ & $\tau \frac{1-\mathcal{I%
}_{H}\left( \tau \right) }{\mathcal{I}_{H}\left( \tau \right) }$ \\ \hline
\end{tabular}
\end{center}

\smallskip

\textbf{Table 2}: Eight pairs of universal inequality criteria that determine the mean performance of sharp restart. For each pair of criteria, the table's columns specify: the inequality index $\mathcal{I}$ on which the criteria are based; the threshold level $l$ that corresponds to the inequality index; and the timer parameters $\tau$ for which the criteria apply. Comparing the inequality index $\mathcal{I}$ to the threshold level $l$, the criteria assert that: if $\mathcal{I}>l$ then mean performance improves; and if $\mathcal{I}<l$ then mean performance worsens. For each pair of criteria -- excluding the CV and Gini criteria -- the table's right column further specifies the critical `mean value' $\mu_{c}$ that is based on the inequality-index value $\mathcal{I}$. Comparing the task's mean completion time $\mu$ to the critical `mean value' $\mu_{c}$, the criteria assert that: if $\mu > \mu_{c}$ then mean performance improves; and if $\mu < \mu_{c}$ then mean performance worsens. Remarks regarding specific inequality indices are the following. Bonferroni: the value $\nu$ is given by Eq. (\ref{325}). Pietra and Vertical: $F\left( \cdot \right)$ and $\bar{F}\left( \cdot \right)$ are, respectively, the distribution function and the survival function of the task's completion time $T$.

\newpage

\textbf{Acknowledgments}. Shlomi Reuveni acknowledges support from the Azrieli Foundation, from the Raymond and Beverly Sackler Center for Computational Molecular and Materials Science at Tel Aviv University, and from the Israel Science Foundation (grant No. 394/19).\\

\section{\label{8}Methods}

\subsection{Derivation of Eq. (\protect\ref{212})}

Introduce the function
\begin{equation}
f_{res}\left( t\right) =\frac{1}{\mu }\bar{F}\left( t\right)  \label{A101}
\end{equation}
($t\geq 0$). As the integral of the input's survival function $\bar{F}\left(t\right) $ is the input's mean, $\mu =\int_{0}^{\infty }\bar{F}\left(t\right) dt$, the function $f_{res}\left( t\right) $ is a probability density: it is non-negative, $f_{res}\left( t\right) \geq 0$; and it has a unit integral, $\int_{0}^{\infty }f_{res}\left( t\right) dt=1$. In effect, $f_{res}\left(t\right) $ is the density function of the input's \textquotedblleft residual lifetime\textquotedblright\ \cite{Ros}. The corresponding distribution and survival functions are $F_{res}\left( t\right) =\int_{0}^{t}f_{res}\left(s\right) ds$ and $\bar{F}_{res}\left( t\right) =\int_{t}^{\infty}f_{res}\left( s\right) ds$. Moreover, the corresponding mean is 
\begin{equation}
\left. 
\begin{array}{l}
\mu _{res}=\int_{0}^{\infty }\bar{F}_{res}\left( t\right)
dt=\int_{0}^{\infty }tf_{res}\left( t\right) dt \\ 
\\ 
=\frac{1}{\mu }\int_{0}^{\infty }t\bar{F}\left( t\right) dt=\frac{1}{2\mu }%
\int_{0}^{\infty }t^{2}f\left( t\right) dt \\ 
\\ 
=\frac{1}{2\mu }\mathbf{E}\left[ T^{2}\right] =\frac{1}{2\mu }\left( \sigma
^{2}+\mu ^{2}\right).
\end{array}
\right.  \label{A102}
\end{equation}
In the second line of Eq. (\ref{A102}) we used integration by parts, and in the third line of Eq. (\ref{A102}) we used the following representation of the input's variance: $\sigma ^{2}=\mathbf{E}\left[ T^{2}\right] -\mathbf{E}%
\left[ T\right] ^{2}$.

Dividing both sides of Eq. (\ref{211}) by $\mu $, and using the distribution function $F_{res}\left( t\right) $, we have
\begin{equation}
\frac{M\left( \tau \right) }{\mu }=\frac{F_{res}\left( \tau \right) }{F\left( \tau \right) }.  \label{A103}
\end{equation}
Eq. (\ref{A103}), together with the coupling between distribution and
survival functions, implies that
\begin{equation}
\left. 
\begin{array}{l}
\frac{M\left( \tau \right) -\mu }{\mu }=\frac{M\left( \tau \right) }{\mu }-1=%
\frac{F_{res}\left( \tau \right) }{F\left( \tau \right) }-1 \\ 
\\ 
=\frac{F_{res}\left( \tau \right) -F\left( \tau \right) }{F\left( \tau
\right) }=\frac{\bar{F}\left( \tau \right) -\bar{F}_{res}\left( \tau \right) 
}{F\left( \tau \right) }.
\end{array}%
\right.  \label{A104}
\end{equation}%
In turn, Eq. (\ref{A104}) implies that
\begin{equation}
\left[ M\left( \tau \right) -\mu \right] F\left( \tau \right) =\mu \left[\bar{F}\left( \tau \right) -\bar{F}_{res}\left( \tau \right) \right].  \label{A105}
\end{equation}
Integrating both sides of Eq. (\ref{A105}), and using Eq. (\ref{A102}), yields Eq. (\ref{212}):
\begin{equation}
\left. 
\begin{array}{l}
\int_{0}^{\infty }\left[ M\left( \tau \right) -\mu \right] F\left( \tau
\right) d\tau =\mu \int_{0}^{\infty }\left[ \bar{F}\left( \tau \right) -\bar{%
F}_{res}\left( \tau \right) \right] d\tau \\ 
\\ 
=\mu \left[ \int_{0}^{\infty }\bar{F}\left( \tau \right) d\tau
-\int_{0}^{\infty }\bar{F}_{res}\left( \tau \right) d\tau \right] =\mu
\left( \mu -\mu _{res}\right) \\ 
\\ 
=\mu ^{2}-\frac{1}{2}\left( \sigma ^{2}+\mu ^{2}\right) =\frac{1}{2}\left(
\mu ^{2}-\sigma ^{2}\right).
\end{array}
\right.  \label{A106}
\end{equation}

\subsection{Derivation of Eqs. (\protect\ref{312}) and (\protect\ref{313})}

The output of the sharp-restart algorithm admits the following stochastic representation \cite{MP1SR}: 
\begin{equation}
T_{R}=\min \left\{ T,\tau \right\} +T_{R}^{\prime }\cdot I\left\{ T>\tau \right\},  \label{A121}
\end{equation}
where $T_{R}^{\prime }$ is an IID copy of the output $T_{R}$. Applying expectation to both sides of Eq. (\ref{A121}) yields
\begin{equation}
\mathbf{E}\left[ T_{R}\right] =\mathbf{E}\left[ \min \left\{ T,\tau \right\} \right] +\mathbf{E}\left[ T_{R}^{\prime }\cdot I\left\{ T>\tau \right\} \right].  \label{A122}
\end{equation}
As $T_{R}^{\prime }$ is an IID copy of $T_{R}$, we have
\begin{equation}
\left. 
\begin{array}{l}
\mathbf{E}\left[ T_{R}^{\prime }\cdot I\left\{ T>\tau \right\} \right] =
\mathbf{E}\left[ T_{R}^{\prime }\right] \cdot \mathbf{E}\left[ I\left\{
T>\tau \right\} \right]  \\ 
\\ 
=\mathbf{E}\left[ T_{R}^{\prime }\right] \Pr \left( T>\tau \right) =\mathbf{E
}\left[ T_{R}\right] \bar{F}\left( \tau \right).
\end{array}
\right.   \label{A123}
\end{equation}
And, substituting Eq. (\ref{A123}) in Eq. (\ref{A122}) we have
\begin{equation}
\mathbf{E}\left[ T_{R}\right] =\mathbf{E}\left[ \min \left\{ T,\tau \right\} 
\right] +\mathbf{E}\left[ T_{R}\right] \bar{F}\left( \tau \right).
\label{A124}
\end{equation}
In turn, Eq. (\ref{A124}) implies that
\begin{equation}
\mathbf{E}\left[ T_{R}\right] F\left( \tau \right) =\mathbf{E}\left[ \min
\left\{ T,\tau \right\} \right],  \label{A125}
\end{equation}
and Eq. (\ref{A125}) yields Eq. (\ref{312}).

In what follows $T_{1}$ and $T_{2}$ are IID copies of the input $T$. The distribution function of the random variable $\max \left\{T_{1},T_{2}\right\} $ is $F\left( t\right) ^{2}$ ($t\geq 0$), and consequently its density function is: $f_{\max }\left( t\right) =2F\left(t\right) f\left( t\right) $ ($t>0$). Multiplying both sides of Eq. (\ref{A125}) by the term $2f\left( \tau \right) $, and using the density function $f_{\max }\left( t\right)$, we have
\begin{equation}
M\left( \tau \right) f_{\max }\left( \tau \right) =2\mathbf{E}\left[ \min
\left\{ T,\tau \right\} \right] f\left( \tau \right).  \label{A131}
\end{equation}
Integrating Eq. (\ref{A131}) yields
\begin{equation}
\left. 
\begin{array}{l}
\int_{0}^{\infty }M\left( \tau \right) f_{\max }\left( \tau \right) d\tau
=2\int_{0}^{\infty }\mathbf{E}\left[ \min \left\{ T,\tau \right\} \right]
f\left( \tau \right) d\tau \\ 
\\ 
=2\int_{0}^{\infty }\left[ \int_{0}^{\infty }\min \left\{ t,\tau \right\}
f\left( t\right) dt\right] f\left( \tau \right) d\tau \\ 
\\ 
=2\int_{0}^{\infty }\int_{0}^{\infty }\min \left\{ t,\tau \right\} f\left(
t\right) f\left( \tau \right) dtd\tau \\ 
\\ 
=2\mathbf{E}\left[ \min \left\{ T_{1},T_{2}\right\} \right].
\end{array}
\right.  \label{A132}
\end{equation}
As $f_{\max }\left( t\right) $ is a density function, Eqs. (\ref{A132}) and (\ref{311}) yield Eq. (\ref{313}):%
\begin{equation}
\left. 
\begin{array}{l}
\int_{0}^{\infty }\left[ \frac{M\left( \tau \right) -\mu }{\mu }\right]
f_{\max }\left( \tau \right) d\tau =\int_{0}^{\infty }\left[ \frac{M\left(
\tau \right) }{\mu }-1\right] f_{\max }\left( \tau \right) d\tau \\ 
\\ 
=\frac{1}{\mu }\int_{0}^{\infty }M\left( \tau \right) f_{\max }\left( \tau
\right) d\tau -\int_{0}^{\infty }f_{\max }\left( \tau \right) d\tau \\ 
\\ 
=\frac{2}{\mu }\mathbf{E}\left[ \min \left\{ T_{1},T_{2}\right\} \right]
-1=2\left( 1-\mathcal{I}_{Gini}\right) -1 \\ 
\\ 
=1-2\mathcal{I}_{Gini}.
\end{array}
\right.  \label{A133}
\end{equation}

\subsection{Derivation of Eqs. (\protect\ref{323}) and (\protect\ref{324})}

Integration by parts implies that%
\begin{equation}
\int_{0}^{\tau }\bar{F}\left( t\right) dt=\int_{0}^{\tau }tf\left( t\right)
dt+\tau \bar{F}\left( \tau \right).  \label{A141}
\end{equation}
Combined together, Eq. (\ref{211}), Eq. (\ref{A141}) and Eq. (\ref{321}) yields Eq. (\ref{323}):
\begin{equation}
\left. 
\begin{array}{l}
M\left( \tau \right) =\frac{1}{F\left( \tau \right) }\int_{0}^{\tau }\bar{F}%
\left( t\right) dt \\ 
\\ 
=\frac{1}{F\left( \tau \right) }\int_{0}^{\tau }tf\left( t\right) dt+\tau 
\frac{\bar{F}\left( \tau \right) }{F\left( \tau \right) } \\ 
\\ 
=\phi \left( \tau \right) +\tau \frac{\bar{F}\left( \tau \right) }{F\left(
\tau \right) }.
\end{array}
\right.  \label{A142}
\end{equation}%
Note that we can write Eq. (\ref{A142}) in the following form%
\begin{equation}
M\left( \tau \right) =\phi \left( \tau \right) +\tau \frac{1}{F\left( \tau
\right) }-\tau. \label{A143}
\end{equation}

Multiplying both sides of Eq. (\ref{A143}) by the term $f\left( \tau \right) 
$ yields%
\begin{equation}
M\left( \tau \right) f\left( \tau \right) =\phi \left( \tau \right) f\left(
\tau \right) +\tau \frac{f\left( \tau \right) }{F\left( \tau \right) }-\tau
f\left( \tau \right).  \label{A144}
\end{equation}
Integrating Eq. (\ref{A144}), and using Eq. (\ref{325}), we have%
\begin{equation}
\left. 
\begin{array}{l}
\int_{0}^{\infty }M\left( \tau \right) f\left( \tau \right) d\tau \\ 
\\ 
=\int_{0}^{\infty }\phi \left( \tau \right) f\left( \tau \right) d\tau
+\int_{0}^{\infty }\tau \frac{f\left( \tau \right) }{F\left( \tau \right) }%
d\tau -\int_{0}^{\infty }\tau f\left( \tau \right) d\tau \\ 
\\ 
=\int_{0}^{\infty }\phi \left( \tau \right) f\left( \tau \right) d\tau +\nu
-\mu.
\end{array}
\right.  \label{A145}
\end{equation}
As $f\left( t\right) $ is a density function, Eqs. (\ref{A145}) and (\ref%
{322}) yield Eq. (\ref{324}):%
\begin{equation}
\left. 
\begin{array}{l}
\int_{0}^{\infty }\left[ \frac{M\left( \tau \right) -\mu }{\mu }\right]
f\left( \tau \right) d\tau =\int_{0}^{\infty }\left[ \frac{M\left( \tau
\right) }{\mu }-1\right] f\left( \tau \right) d\tau \\ 
\\ 
=\frac{1}{\mu }\int_{0}^{\infty }M\left( \tau \right) f\left( \tau \right)
d\tau -\int_{0}^{\infty }f\left( \tau \right) d\tau \\ 
\\ 
=\frac{1}{\mu }\left[ \int_{0}^{\infty }\phi \left( \tau \right) f\left(
\tau \right) d\tau +\nu -\mu \right] -1 \\ 
\\ 
=\frac{1}{\mu }\int_{0}^{\infty }\phi \left( \tau \right) f\left( \tau
\right) d\tau +\frac{\nu }{\mu }-2 \\ 
\\ 
=\left( 1-\mathcal{I}_{Bonf}\right) +\frac{\nu }{\mu }-2=\frac{\nu -\mu }{%
\mu }-\mathcal{I}_{Bonf}.
\end{array}
\right.  \label{A146}
\end{equation}

\subsection{Derivation of Eq. (\protect\ref{400})}

Note that%
\begin{equation}
\max \left\{ T,\tau \right\} +\min \left\{ T,\tau \right\} =T+\tau,
\label{A151}
\end{equation}%
and%
\begin{equation}
\max \left\{ T,\tau \right\} -\min \left\{ T,\tau \right\} =\left\vert
T-\tau \right\vert.  \label{A152}
\end{equation}%
Subtracting Eq. (\ref{A152}) from Eq. (\ref{A151}) yields%
\begin{equation}
2\min \left\{ T,\tau \right\} =T+\tau -\left\vert T-\tau \right\vert. \label{A153}
\end{equation}%
In turn, applying expectation to both sides of Eq. (\ref{A153}) we have%
\begin{equation}
\left. 
\begin{array}{l}
2\mathbf{E}\left[ \min \left\{ T,\tau \right\} \right] =\mathbf{E}\left[
T+\tau -\left\vert T-\tau \right\vert \right] \\ 
\\ 
=\mathbf{E}\left[ T\right] +\mathbf{E}\left[ \tau \right] -\mathbf{E}\left[
\left\vert T-\tau \right\vert \right] \\ 
\\ 
=\mu +\tau -\mathbf{E}\left[ \left\vert T-\tau \right\vert \right].
\end{array}
\right.  \label{A154}
\end{equation}
Combining together Eq. (\ref{312}) and Eq. (\ref{A154}) yields Eq. (\ref{400}
):
\begin{equation}
M\left( \tau \right) =\frac{\mathbf{E}\left[ \min \left\{ T,\tau \right\} %
\right] }{F\left( \tau \right) }=\frac{\mu +\tau -\mathbf{E}\left[
\left\vert T-\tau \right\vert \right] }{2F\left( \tau \right) }.
\label{A155}
\end{equation}

\subsection{Derivation of Eqs. (\protect\ref{511}) and (\protect\ref{521})}

Combining together Eq. (\ref{A142}) and Eq. (\ref{501}) we have
\begin{equation}
\left. 
\begin{array}{l}
M\left( \tau \right) =\frac{1}{F\left( \tau \right) }\int_{0}^{\tau
}tf\left( t\right) dt+\tau \frac{\bar{F}\left( \tau \right) }{F\left( \tau
\right) } \\ 
\\ 
=\mu \frac{F_{dol}\left( \tau \right) }{F\left( \tau \right) }+\tau \frac{%
\bar{F}\left( \tau \right) }{F\left( \tau \right) }.
\end{array}
\right.  \label{A161}
\end{equation}
Eq. (\ref{A161}) yields Eq. (\ref{511}):
\begin{equation}
\left. 
\begin{array}{l}
\frac{M\left( \tau \right) -\mu }{\mu }=\frac{1}{\mu }M\left( \tau \right) -1
\\ 
\\ 
=\frac{1}{\mu }\left[ \mu \frac{F_{dol}\left( \tau \right) }{F\left( \tau
\right) }+\tau \frac{\bar{F}\left( \tau \right) }{F\left( \tau \right) }%
\right] -1 \\ 
\\ 
=\frac{\tau }{\mu }\frac{\bar{F}\left( \tau \right) }{F\left( \tau \right) }-%
\left[ 1-\frac{F_{dol}\left( \tau \right) }{F\left( \tau \right) }\right].
\end{array}%
\right.  \label{A162}
\end{equation}

Eq. (\ref{A162}) implies that
\begin{equation}
\left. 
\begin{array}{l}
\left[ M\left( \tau \right) -\mu \right] F\left( \tau \right) =\tau \left[
1-F\left( \tau \right) \right] -\mu \left[ F\left( \tau \right)
-F_{dol}\left( \tau \right) \right] \\ 
\\ 
=\tau \bar{F}\left( \tau \right) -\mu \left[ \bar{F}_{dol}\left( \tau
\right) -\bar{F}\left( \tau \right) \right] =\left( \tau +\mu \right) \bar{F}%
\left( \tau \right) -\mu \bar{F}_{dol}\left( \tau \right).
\end{array}
\right.  \label{A163}
\end{equation}
Eq. (\ref{A163}) yields
\begin{equation}
\left. 
\begin{array}{l}
\frac{M\left( \tau \right) -\mu }{\mu }\left[ \frac{\mu }{\tau +\mu }\frac{%
F\left( \tau \right) }{\bar{F}_{dol}\left( \tau \right) }\right] =\frac{%
M\left( \tau \right) -\mu }{\tau +\mu }\frac{F\left( \tau \right) }{\bar{F}%
_{dol}\left( \tau \right) } \\ 
\\ 
=\frac{1}{\tau +\mu }\frac{\left( \tau +\mu \right) \bar{F}\left( \tau
\right) -\mu \bar{F}_{dol}\left( \tau \right) }{\bar{F}_{dol}\left( \tau
\right) }=\frac{\bar{F}\left( \tau \right) }{\bar{F}_{dol}\left( \tau
\right) }-\frac{\mu }{\tau +\mu } \\ 
\\ 
=\left[ 1-\frac{\mu }{\tau +\mu }\right] -\left[ 1-\frac{\bar{F}\left( \tau
\right) }{\bar{F}_{dol}\left( \tau \right) }\right] \\ 
\\ 
=\frac{\tau }{\tau +\mu }-\left[ 1-\frac{\bar{F}\left( \tau \right) }{\bar{F}
_{dol}\left( \tau \right) }\right].
\end{array}
\right.  \label{A164}
\end{equation}
In turn, Eq. (\ref{A164}) yields Eq. (\ref{521}).

\end{document}